\documentclass[useAMS,usenatbib]{mn2e}
\usepackage{amsmath}
\usepackage{amssymb}
\usepackage{graphicx}
\usepackage{euscript}
\usepackage{bm,tabularx}
\usepackage{float}
\usepackage[usenames,dvipsnames]{pstricks}
\usepackage{epsfig}
\usepackage{pst-grad} 
\usepackage{pst-plot} 
\usepackage{lipsum, mathtools}
\usepackage{comment}
\usepackage{enumerate}
\usepackage{wrapfig,multicol}
\usepackage{color,colortbl}
\usepackage[colorlinks,citecolor=blue,linkcolor=red,urlcolor=blue]{hyperref}
\graphicspath{{./figures/}}

\newcommand{\beq}{\begin{equation}}
\newcommand{\eeq}{\end{equation}}

\newcommand{\bfa}{\mbox{\boldmath $a$}}

\newcommand{\bfc}{\mbox{\boldmath $c$}}

\newcommand{\bfk}{\mbox{\boldmath $k$}}

\newcommand{\bfq}{\mbox{\boldmath $q$}}
\newcommand{\bfu}{\mbox{\boldmath $u$}}
\newcommand{\bfv}{\mbox{\boldmath $v$}}
\newcommand{\bfx}{\mbox{\boldmath $x$}}

\newcommand{\bfA}{\mbox{\boldmath $A$}}
\newcommand{\bfB}{\mbox{\boldmath $B$}}
\newcommand{\bfC}{\mbox{\boldmath $C$}}

\newcommand{\bfH}{\mbox{\boldmath $H$}}
\newcommand{\bfK}{\mbox{\boldmath $K$}}

\newcommand{\bfS}{\mbox{\boldmath $S$}}
\newcommand{\bfU}{\mbox{\boldmath $U$}}

\newcommand{\bfX}{\mbox{\boldmath $X$}}

\newcommand{\bfQ}{\mbox{\boldmath $Q$}}

\newcommand{\ex}{\mbox{{\boldmath $e$}}_{1}}
\newcommand{\ey}{\mbox{{\boldmath $e$}}_{2}}
\newcommand{\ez}{\mbox{{\boldmath $e$}}_{3}}

\newcommand{\bnabla}{\mbox{\boldmath $\nabla$}}
\newcommand{\cross}{\mbox{\boldmath $\times$}}
\newcommand{\cendot}{\mbox{\boldmath $\cdot\,$}}
\newcommand{\rem}{{\rm Rm}}
\newcommand{\re}{{\rm Re}}

\newcommand{\ud}{{\rm d}}
\newcommand{\bsy}{\boldsymbol}
\newcommand{\Eq}[1]{Eq.~(\ref{#1})}

\newcommand{\Fig}[1]{Fig.~\ref{#1}}
\newcommand{\Sec}[1]{Section~(\ref{#1})}

\def\half{{\textstyle{1\over 2}}}
\definecolor{dred}{rgb}{.9, .15,.3}
\definecolor{brown}{rgb}{0.5,0.2,0.2}
\definecolor{dgreen}{rgb}{0.15,0.6,0.15}
\definecolor{dblue}{rgb}{0.0,0.0,0.6}

\definecolor{midblue}{rgb}{0.0,0.4,0.7}
\definecolor{midgreen}{rgb}{0.0,0.8,0.3}

\begin{document}

\title[Stochastic helicity shear dynamo]
{Mean field dynamo action in shearing flows. II:
fluctuating kinetic helicity with zero mean}

\author[Jingade \& Singh]
{Naveen Jingade$^{1,2}$\thanks{E-mail: naveen@rri.res.in}
and Nishant K. Singh$^{3,4}$\thanks{E-mail: nishant@iucaa.in}\\
 $^{1}$Indian Institute of Science, Bangalore 560 012, India\\
 $^{2}$Raman Research Institute, Sadashivanagar, Bangalore 560 080, India\\
 $^{3}$Inter-University Centre for Astronomy \& Astrophysics, Post Bag 4, Ganeshkhind, Pune 411 007, India\\
 $^{4}$Max Planck Institute for Solar System Research, Justus-von-Liebig-Weg 3,
 37707 G\"ottingen, Germany \\
}

\pagerange{\pageref{firstpage}--\pageref{lastpage}} \pubyear{}

\maketitle

\label{firstpage}

\begin{abstract}
Here we explore the role of temporal fluctuations in kinetic helicity on the generation of
large-scale magnetic fields in presence of a background linear shear flow.
Key techniques involved here are same as in our earlier work \citep[][hereafter paper~I]{JS20},
where we have used the renovating flow based model with shearing waves. Both, the velocity and the
helicity fields, are treated as stochastic variables with finite correlation times,
$\tau$ and $\tau_h$, respectively. Growing solutions are obtained when $\tau_h > \tau$, even
when this time-scale separation, characterised by $m=\tau_h/\tau$, remains
below the threshold for causing the turbulent
diffusion to turn negative. In regimes when turbulent diffusion remains positive,
and $\tau$ is on the order of eddy turnover time $T$, the axisymmetric modes display
non-monotonic behaviour with shear rate $S$: both, the growth rate $\gamma$ and
the wavenumber $k_\ast$ corresponding to the fastest growing mode, first increase, reach
a maximum and then decrease with $|S|$, with $k_\ast$ being always smaller than
eddy-wavenumber, thus boosting growth of magnetic fields at large length scales.
The cycle period $P_{\rm cyc}$ of growing dynamo wave is inversely proportional to $|S|$
at small shear, exactly as in the fixed kinetic helicity case of paper~I. This dependence
becomes shallower at larger shear. Interestingly enough, various curves corresponding to
different choices of $m$ collapse on top of each other in a plot of $m P_{\rm cyc}$ with
$|S|$.
\end{abstract}



\section{Introduction}

Magnetic fields are hosted by almost all astrophysical
objects like Sun, galaxies, accretion disks, etc \citep{Par79,RSS88,JL17}. All such objects
have conducting plasma that are turbulent in nature.
Magnetic fields are expected to be dissipated by the turbulent diffusion, therefore, continuous magnetic field generation is needed to maintain them.
The process by which a seed magnetic field is amplified by continuous conversion
of kinetic energy of the plasma into the magnetic energy, without currents at infinity,
is universally known as the dynamo action.
A popular paradigm for the generation of magnetic field at large spatial scales
involves the presence of net kinetic helicity in the medium, leading to the $\alpha$-effect \citep{Mof78,KR80,BS05}.

Large-scale dynamo (LSD) is observed to operate in shearing-box simulations with
non-helical driving \citep{BRRK08,You08a,SJ15}. In such setups that lead to the
shear dynamo, the
usual $\alpha$-effect is absent, although it can fluctuate in time
\citep[see][for details]{BRRK08}.
This topic of shear dynamo has generated much interest
\citep{kleeorin2003,RK08,SS09a,SS09b,SS10,SS11} and it is still somewhat an open problem
\citep[see][also for various related issues, and references therein]{Kap20}.
Possibility of a coherent LSD driven
by a combination of strong magnetic fluctuations and shear has also been proposed \citep{SB15b}.
It is noted in \cite{You08b} that the shear dynamo operates well below the threshold of
fluctuation dynamo, therefore suggesting that the underlying cause of dynamo must be a kinematic
process.
Emerging wisdom seems to indicate that the alpha-fluctuations, coupled with shear, could
cause a shear dynamo, also known as incoherent alpha effect
\citep{VB97,BRRK08,HMS11,MB12,SS14,JNS18,Kap20}. 
Our interest in this and other related earlier works has been on the kinematic problem
when initial seed magnetic fields are weak.

\begin{figure}
\includegraphics[scale=0.28]{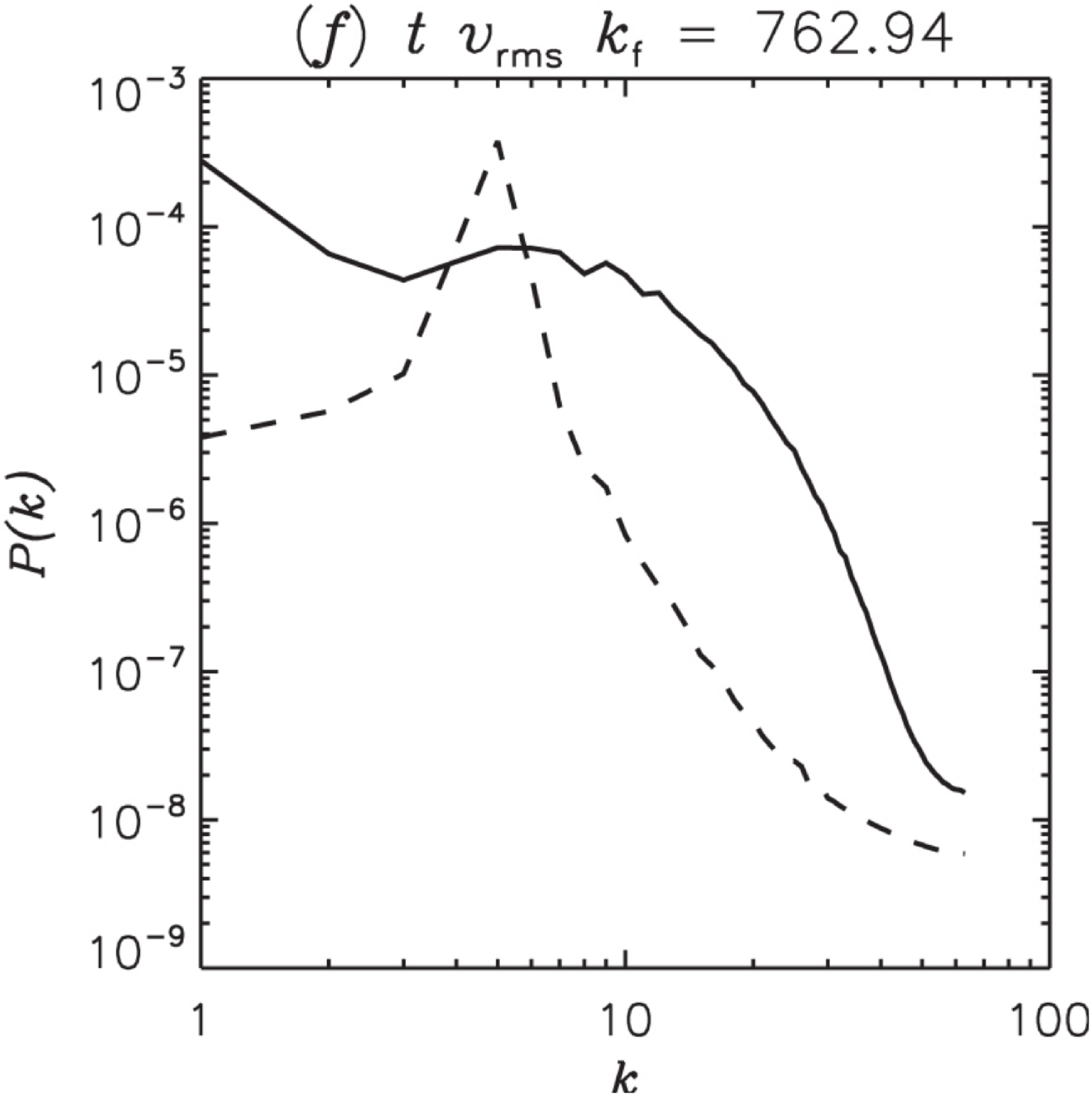}
\caption{Kinetic (dashed) and magnetic (solid) energy power spectra for $\re=0.641$ and
$\rem=32.039$, the forcing wavenumber $k_f/K = 5.09$ and shear rate $S_h=S/v_{\rm rms}k_f=-0.6$
from the time when the magnetic energy is saturated; based on a typical numerical setup
of the shear dynamo. Taken from \citet{SJ15} \copyright AAS, and reproduced here with permission.}
\label{pow}
\end{figure}

\citet{SS10,SS11} showed that the shear dynamo is not possible when the kinetic
helicity vanishes point-wise.
To understand the importance of non-linearity in the Navier-Stokes equation for the shear dynamo,
\cite{SJ15} explored the regime $\re<1$ and $\rem>1$, and showed that the shear
dynamo operates well in this case too.
Figure \ref{pow} illustrates power spectra of kinetic and magnetic energies in the saturated
state of the shear dynamo at low $\re$ \citep[adapted from][]{SJ15}. Kinetic power
is predominantly concentrated at single scale, the forcing scale $k_f^{-1}$, and
thus the flow is nearly monochromatic. This
indicates that the flow non-linearity may not be a critical ingredient for the shear dynamo,
and we could pursue this problem by just considering a single scale flow.
Another advantage of this regime is that, due to heavy viscous dissipation, it leads to
the suppression of vorticity dynamo \citep{EKR03,KMB09} which may also be excited
in these experiments, causing, in a sense, unnecessary complications in our attempt to understand
the underlying mechanism of the shear dynamo. In real systems, rotation suppresses
the vorticity dynamo \citep{You08b}. Based on such outcomes from these studies, we aim
to build a minimalistic model of shear dynamo.

In this paper, we focus on the effect of zero-mean helicity fluctuations in time,
together with a linear shear, on the generation of large-scale magnetic field.
Without shear, \cite{Kra76} studied the possibility of LSD, driven solely by helicity
fluctuations in time. He conjectured there that the helicity can fluctuate over an intermediate
scale which is larger than turbulent scale of the flow and smaller than the scale of
magnetic field, and termed such a turbulence as \emph{non-normal}. In a double-averaging
scheme \cite[see also][for renormalisation group technique]{Mof83} as proposed in \cite{Kra76}, the fluctuations in helicity lead to $\alpha$-fluctuations,
which cause a decrement in the turbulent diffusion. Sufficiently strong $\alpha$ fluctuations
could lead to LSD by the process of negative turbulent diffusion; see \cite{JNS18} for a
generalized model that also includes memory effects as well as the spatial fluctuations
in $\alpha$ leading to the Moffatt drift \citep{Mof78}. This has found applications
in the studies of objects such as, the Sun \citep{Sil00},
accretion disks \citep{VB97}, and galaxies \citep{Sok97}.

Numerical experiments in \citet{You08a,BRRK08}, and measurements of fluctuations in
all components of tensorial $\alpha_{ij}$ \citep{BRRK08} motivated a number of studies
\citep{HMS11,MB12,RP12,McW12,SS14}, which explored the effects of $\alpha$-fluctuations
in shearing background. Most of these studies found the possibility of growth in the
second moment of the mean magnetic
field, i.e., mean magnetic energy. First moment could grow only as a result of negative
turbulent diffusion caused by strong $\alpha$ fluctuations \citep{Kra76,MB12,JNS18}.
Net negative diffusion is not observed in the simulations of \cite{BRRK08} in the range of
parameters explored where shear dynamo was found to operate.
Earlier analytical models were further generalized in \citet{SS14,JNS18} to also include
the memory effects, to investigate whether the LSD would be possible when temporal
$\alpha$ fluctuations were weak, i.e., when overall diffusion remains positive.
In this more appropriate regime, no dynamo action was seen in case of purely temporal $\alpha$-fluctuations. However, a new idea emerged in these studies, in terms of the Moffatt
drift, which would be finite in case of statistically anisotropic $\alpha$ fluctuations.
With finite memory, this alone could drive an LSD \citep{S16}, and the role
of shear is to further aid the dynamo action \citep{SS14,JNS18}.
Nevertheless, the challenge to explain the growth of mean-magnetic field, the first moment,
in a simpler setting with purely temporal $\alpha$ fluctuations, remained.

One possible caveat in these analytical approaches could be the assumption
of linear algebraic relation between electromotive force (EMF) and the transport
coefficients\footnote{This is an assumption of local and instantaneous
interaction of the mean magnetic field with the EMF.}.
This is a serious limitation, which we remedy here by starting the analysis from
the magnetic induction equation, which embodies the effect of non-local transfer of energy
through $(\bfB\cendot\bnabla)\bfv$ term, as we have carried out in paper I. We
found non-trivial signature of non-local interaction in the cycle period of the dynamo wave.
Another limitation in previous works could be that the $\alpha$ itself
is assumed to be unaffected by shear. We overcome this in the present work by
considering plane shearing waves, that are time-dependent exact solutions to the Navier-Stokes
equations \citep{SS17}, allowing us to self-consistently include the anisotropic
effect of shear on the stochastic flow. We follow the techniques of paper~I and
extend the renovating model to include the helicity fluctuations;
see also \cite{GB92,KSS12}, for more details on the renovating flow model.
Thus, instead of $\alpha$ fluctuations, here we directly consider fluctuations in the kinetic
helicity of the stochastic flow.
We let the helicity fluctuations to have the renovation time ($\tau_h$) different from the
velocity renovation time $\tau$. In the limit of infinitely long $\tau_h$, the present
work reduces to the work of paper~I which focussed on the case of fixed kinetic helicity. 

In \textsection~\,\ref{Mod}, we briefly review the shearing plane waves solutions to the
Navier-Stokes equation; details are given in \cite{SS17} and paper~I.
We derive the expression for the general response tensor for the mean-magnetic field
incorporating the helicity fluctuations in \textsection~\,\ref{EMF}.
In \textsection~\,\ref{HF}, we derive a condition for the growth rate of the
mean-magnetic field to be positive in the absence of shear. In the short time-correlation limit, we show
that the dispersion relation obtained from the response tensor of the magnetic field
resembles the result derived for $\alpha$-fluctuations given in \cite{Kra76}. Then we present
the result for the case of helicity fluctuations in the presence of shear in
\textsection~\,\ref{HFWS}. Growing solutions for the mean-magnetic fields are obtained
even without the need for any net negative turbulent diffusion. In \textsection~\,\ref{Diff},
we discuss about the two different origins for the fluctuations in $\alpha$. Implications
of our results for earlier numerical findings on the shear dynamo is also discussed.
We conclude in \textsection~\,\ref{Conclude}.

\section{Model system}
First we formulate the problem to investigate the evolution of the mean-magnetic field in
the background shear flow with the turbulence having fluctuating helicity. We consider ($\ex,\ey,\ez$) as the orthonormal unit vectors in the Cartesian coordinate system in the lab frame, where $\bfX=(X_1,X_2,X_3)$ is considered as position vector.
The mean shear is chosen to act along the $\ey$ direction, varying linearly with $X_1$. This is a local shearing--sheet approximation to the differential
rotating  disks \citep{GL65,BRRK08}. The model velocity field $\bfU$ in such an approximation can be written as, $\bfU(\bfX,t) = S\,X_1\ey + \bfu(\bfX,t)$, where $\bfu$ is the turbulent velocity field, and the shear rate $S$ is a constant parameter.

\subsection{Renewing flows in shearing background}  
\label{Mod}
The inviscid Navier--Stokes (NS) equation with for the model velocity field
$\bfU$ is
\begin{equation}
\left(\frac{\partial}{\partial t}+SX_1\frac{\partial}{\partial X_2}\right)\bfu+Su_1\ey +(\bfu\cendot\bnabla)\bfu=-\bnabla p\,.
\label{NSeqn}
\end{equation}
Here $\bnabla\cendot\bfu = 0$, as we consider the flow $\bfu$ to be incompressible,
for simplicity.
The single helical wave solution for NS-equation is of the following form,  
\begin{align}
\bfu(\bfX,t) = & \bfA(St,\bfq)\sin(\bfQ(t)\cendot\bfX + \Psi) \nonumber \\ & + h\,\bfC(St, \bfq)\cos(\bfQ(t)\cendot\bfX + \Psi)
\label{vel}
\end{align}
where $\bfQ(t)$ is a shearing wavevector and its form is given by $\bfQ = (q_1-S q_2(t-t_0),q_2,q_3)$; $\bfq = (q_1,q_2,q_3)$ is the wavevector at initial time $t_0$, $\Psi$ is the phase of the wave, $\bfA(St,\bfq)$ and
$\bfC(St,\bfq)$ are the amplitudes of the sheared helical wave, with $\bfa$ and $\bfc$
denoting their initial values, respectively.
This shearing form of the wave vector arises because of the inhomogeneity of the \Eq{NSeqn}
in the variable $X_1$.
When we substitute \Eq{vel} in (\ref{NSeqn}), the differential equations for the amplitudes are obtained, which can be solved to obtain the expression for the amplitudes; see paper~I and
\cite{SS17} for further details.
The value of $h$ controls the helicity of the flow $\bfu$
and it varies in the range $[-1,1]$.
The incompressibility condition yields the following: 
$\bfQ(t)\cendot\bfA(t) = 0$ and $\bfQ(t)\cendot\bfC(t) = 0$.

\begin{figure}
\includegraphics[scale=0.4]{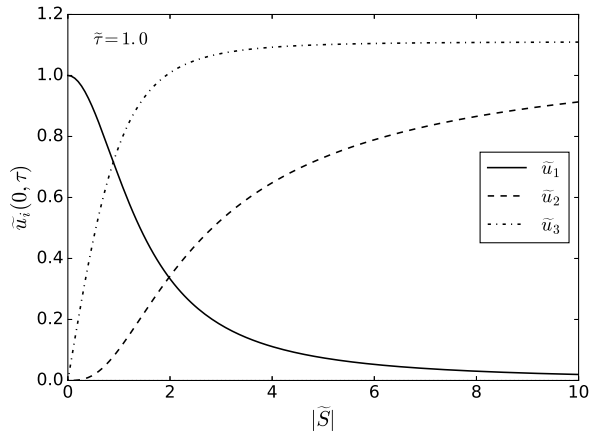}
\caption{Plot of velocity amplitudes as functions of shear/time. The initial wavevector
and amplitudes are $\bfq = (0,-1,1)$ and $\bfa = (1,0,0)$, respectively,
and the phase $\Psi$ is chosen to be zero.}
\label{vel_amp}
\end{figure}

The amplitudes are sheared in the opposite sense compared to the shearing of the wavevector $\bfQ(t)$ to maintain the incompressibility. This leads to the constancy of phase of the
wave, i.e., $\bfQ(t)\cendot\bfX = \bfq\cendot\bfx_0$, where $\bfx_0$ is initial position
of the fluid particle at time $t_0$. Thus, we can easily integrate
\Eq{vel} to obtain the Lagragian trajectory of the fluid particle.
This is used in the Cauchy's solution given below in \Eq{Cauchy}.
These single scale flows are exact solution to the NS-equation and they may be used
in the study of turbulent dynamos.
Previously, single scale flows have been considered to study different aspects of the
dynamo \citep{HY08, TC08}.
Here we aim to understand the generation of large-scale magnetic field in a scenario
when the stochastic component of the flow is itself affected by the background shear.

We give below the expression for the most relevant component of the amplitude:
\beq
A_1(\bfq, t) \;=\; \frac{q^2}{\left(q_1 - St q_2\right)^2 + q_2^2 + q_3^2}\,
A_1(\bfq, 0)\,,
\label{amp1}
\eeq
which has an implication for the dynamo at strong shear. For the properties of the other
components we refer the reader to paper~I and \cite{SS17}. 

To understand the effect of shear on the flow $\bfu$, it is convenient to plot the velocity
field at the origin $\bfX = (0,0,0)$ and with phase $\Psi = 0$. The initial amplitudes
$\bfu({\bf 0},0)=\bfa$ and initial wavevector $\bfq$ are chosen to be $(1,0,0)$ and
$(0,-1,1)$, respectively. In Figure \ref{vel_amp}, it is shown that the normalised velocity field
$u_1$ corresponding to the amplitude $A_1$ decays as shear is increased
(as can be seen from \Eq{amp1} also), whereas the velocity field components $u_2$ and $u_3$
become constant with shear. The background flow shears out the $u_1$ component, and generates
the $u_2$ and $u_3$ component. Thus, strong shear creates quasi-two dimensional
flow\footnote{At least at the larger scale, which are affected by the shear, whereas velocity
field at smaller scales may not be much affected by the shear.}. Its implications are
explored in later sections. 

We now construct the random renewing flow ensemble from the shearing waves. In renovating
flow, the time is split into equal interval of length $\tau$ and the velocity field is
distributed randomly and statistically independent across the intervals. And this velocity
field is drawn from the underlying probability distribution function (PDF). The ensemble is
chosen such that: $\Psi$ is distributed uniformly over the interval $[0,2\pi]$; the initial
wavevector $\bfq$ is distributed randomly over the sphere of radius $q$ and the initial
amplitudes $\bfa$ and $\bfc$ are perpendicular to $\bfq$, are therefore distributed randomly
over the circles of radii $a$ and $c$, respectively;
see paper I and \cite{GB92}, for details.

\subsection{Evolution of mean-magnetic field: helicity fluctuations}
\label{EMF}
The evolution of the magnetic field in the conducting plasma is governed by the induction
equation. With total velocity field as $\bfU = SX_1\ey + \bfu$, the evolution equation for the
total magnetic field is,
\beq
\left(\frac{\partial}{\partial t} + SX_1\frac{\partial }{\partial X_2}\right)\bfB +(\bfu\cendot\bnabla)\bfB-S\,B_1 \ey = (\bfB\cendot\bnabla)\bfu + \eta\nabla^2 \bfB
\label{InducSh1}
\eeq
Our interest is in the evolution of the mean-magnetic field. In astrophysical scenarios, the
plasma is highly conducting, leading to very small $\eta$, i.e., $\eta\to 0$. We
also noted in paper~I \citep[see also][section 11.4]{DUO93,CG95}, that in high conductivity
limit, neglecting the diffusion term in the \Eq{InducSh1} has negligible effect on the growth
rate of the mean-magnetic field. Therefore we omit the $\eta$-term in the following to keep the
analysis simple.

Since \Eq{InducSh1} is inhomogeneous in the $X_1$ co-ordinate, employing shearing transformation \citep{SS09b,SS10} is more suitable, 
\beq
X_i = \gamma_{ij}x_j, \quad \mbox{where}\;\;\gamma_{ij} = \delta_{ij} + S\,(t-t_0)\,\delta_{i2}\delta_{j1}\,. 
\label{shear-Trans}
\eeq
Here $\bfx$ and $t_0$ represents the Lagrangian coordinate of the fluid element carried along by background mean shear flow and the initial time, respectively.
Let us rewrite \Eq{InducSh1} in the coordinate $\bfx$ and the difference in time $s=t-t_0$. Introducing new vector functions $\bfH(\bfx,s) = \bfB(\bfX,t)$, $\bfv(\bfx,s)=\bfu(\bfX,t)$, we can write
\begin{align}
&\frac{\partial \bfH}{\partial s}  +(\bfv\cendot\bnabla)\bfH-S\,H_1 \ey = (\bfH\cendot\bnabla)\bfv,\nonumber \\[1em]
& \quad\mbox{with} \quad \bnabla\cendot\bfv = 0, \quad  \bnabla\cendot\bfH = 0\, ,
\label{InducSh2}
\end{align}
\beq
\mbox{where} \quad \bnabla = \frac{\partial}{\partial \bfx}- S\,s\, \ex\frac{\partial}{\partial x_2} \quad \text{is a time dependent operator.} \nonumber
\eeq
 The Cauchy solution to the above equation can be written as  
 \beq
H_i(\bfx, t) = \gamma_{ij}(s){\cal J}_{jk}(\bfx_0,s)H_{k}(\bfx_0,t_0)
\label{Cauchy}
\eeq
where we need to substitute for $\bfx_0$, the initial position of the particle in terms of its final position $\bfx$ by inverting the trajectory $\bfx = \bfx(\bfx_0,s)$. The Jacobian is given by
(see paper~I for further details)
\beq
{\cal J}_{ij} =  \frac{\partial x_i}{\partial x_{0j}} = \delta_{ij}+\displaystyle \int_0^s \left(\frac{\partial v_i}{\partial x_{0j}}-S s'\delta_{i2}\frac{\partial v_1}{\partial x_{0j}}\right)\ud s'\,.
\label{Jacobian}
\eeq
 We can write the evolution of the magnetic field in the general interval $[(n-1)\tau,n\tau]$ as, 
\beq
H_i(\bfx, n\tau) = \gamma_{ij}(\tau){\cal J}_{jk}(\bfx_0,\tau)H_{k}(\bfx_0,(n-1)\tau)
\label{Finalcauchy}
\eeq
owing to the fact that velocity fields are assumed to be independent in each renovating interval. Here we have considered $t_0$ and $t$ as the time $(n-1)\tau$ and $n\tau$, respectively for each interval. 
We define the Fourier transform for the average of magnetic field $\left\langle H_i(\bfx, n\tau)\right\rangle$ in terms of the shearing waves as
\beq
\widetilde{H}_i(\bfk,n\tau) = \int \left\langle H_i(\bfx, t)\right\rangle \exp[-i\bfK_{n-1}\cendot\bfx]\,\ud^3 x
\label{Fou-Trans2}
\eeq
where $\bfK_{n-1} = [k_1-(n-1)S\tau k_2,k_2,k_3]$ is the initial wavevector at time $(n-1)\tau$ (see \cite{SS10,SS11} and paper~I for details) and $\bfk$ is the wavevector at time $t=0$, i.e.,
when $\bfk=\bfK_0$. 
We obtain the Fourier mode of the magnetic field by performing the average over the initial randomness of the magnetic field and statistical ensemble of the velocity field, 
\beq
\widetilde{H}_i(\bfk,n\tau) = {\cal G}_{ik}(\bfK_{n-1},\tau)\widetilde{H}_{k}(\bfk,(n-1)\tau),
\label{MatrixEq}
\eeq
where
\beq
{\cal G}_{ik}(\bfK_{n-1},\tau) = \left\langle \gamma_{ij}(\tau){\cal J}_{jk}(\bfx_0,\tau)e^{-i\,\bfK_{n-1}\cendot(\bfx_0-\bfx)} \right\rangle
\label{res-tens}
\eeq
is the response tensor for the evolution of mean magnetic field.
Since the velocity field statistics are homogeneous in the shearing coordinate $\bfx$, the averaging and the response tensor for the magnetic field becomes independent of the spatial coordinates.   

Note that, in paper~I we have averaged over the velocity statistics of the flow with helicity $h$
held fixed all throughout. But here we attempt to understand the results of \cite{You08b,BRRK08},
where they find mean field dynamo due to non-helical turbulence in shear flow, and therefore,
we allow helicity to fluctuate in time with zero mean such that it takes random and independent value in successive intervals of time.
Helicity fluctuations are expected to play critical role in driving the shear dynamo.
PDFs of $\alpha_{ij}$ as measured in \cite{BRRK08} suggest that the point-wise kinetic helicity
does not vanish even though the driving is non-helical. The fact that the shear dynamo
operates well below the threshold of fluctuation dynamo \citep{BRRK08,You08b} indicates that the
process responsible for the growth of mean magnetic field must be kinematic in nature. Earlier
theoretical studies on the shear dynamo \citep{HMS11,MB12,JNS18} find growing solutions for
the mean magnetic field,
i.e., the first moment, only by the process of negative turbulent diffusion -- a conclusion similar
to that in \cite{Kra76}, but such a negative diffusion is not observed in the numerical studies
\citep{BRRK08,DP09}. The challenge remained therefore to explain the dynamo mechanism when
diffusion remain positive. We demonstrate this possibility in this work.
   
Time-scales of variations of fluctuating velocity and kinetic helicity fields need not be
same \citep{Kra76,Sok97}. Let us assume, in our case, the correlation time of helicity fluctuations
is greater than the correlation time of velocity field fluctuations. In renovating flow model
being adopted here, the durations of time intervals are fixed, and we consider renovation time of
helicity to be an integral multiple of the renovation time of the velocity field, i.e,
$\tau_h=m\tau$, where $\tau$ is the velocity renovation time, which is same as correlation time of
the flow, $\tau_h$ is the renovation time of helicity and $m\geq 2$. 
 
Now we will derive the expression for the response tensor incorporating the helicity fluctuations.
For the illustration purpose, let us consider $m=2$ case,
we have already obtained the response tensor in \Eq{res-tens} by averaging over the statistics of
the velocity field with fixed $h$. Next, we multiply two copies of those response tensors from the
adjacent intervals of time and the final product is averaged over the statistics of the stochastic
helicity. Mathematically, we can relate the magnetic field at $n\tau$ to its value at $(n-2)\tau$,
in general, $(n-m)\tau$, by employing the double averaging scheme as,  
\beq
\hat{H}_i(\bsy{k},n\tau)=\left\langle {\cal G}_{ij}(\bfK_{n-1},\tau){\cal G}_{jk}(\bfK_{n-2},\tau)\right\rangle_{h} \hat{H}_k(\bsy{k},(n-2)\tau)\nonumber\\
\label{magnetic_evolution}
\eeq
where $\bfK_{n} = (k_1-n\,S\tau\,k_2,k_2,k_3)$. 
In general if $\tau_h=m\tau$, we can write magnetic field at $n\tau$ in terms of the magnetic field at $(n-m)\tau$ as,
\begin{align}
&\hat{H}_i(\bsy{k},n\tau)= G_{il}(\bfk,n,m,\tau)\hat{H}_l(\bsy{k},(n-m)\tau)\nonumber \\[1em]
& \mbox{where} \nonumber \\[1em]
& G_{il}(\bfk,n,m,\tau)=\left\langle {\cal G}_{ij}(\bfK_{n-1},\tau)\dots{\cal G}_{kl}(\bfK_{n-m},\tau)\right\rangle_{h}\,.
\label{tensorG}
\end{align}
The magnetic fields will grow if the absolute value of leading eigenvalue of the double-averaged
response tensor $\hat{G}$ ($\equiv [G_{il}]$) is greater than unity. The final magnetic field will
be the eigenvector corresponding to the leading eigenvalue of the response tensor. If $\sigma$ is
the leading eigenvalue, then the exponential growing exponent is defined as,
\beq
\lambda = \frac{1}{m\tau}\ln \sigma(\bfk,n,m,\tau) = \frac{1}{m\tau}\ln|\sigma| + i \frac{{\rm arg}(\sigma)}{m\tau}\,.
\eeq
The real part of the exponent defines the growth rate of the magnetic field, and the cycle period of
the dynamo wave is given by the imaginary part of the exponent:
\beq
\gamma = \frac{1}{m\tau}\ln |\sigma(\bfk,n,m,\tau)|\,; \qquad P_{\rm cyc} = \frac{2\pi m \tau}{{\rm arg}(\sigma)}. 
\label{gam_cyc}
\eeq 

\section{helicity fluctuation without shear}
\label{HF}
In this section, we study the helicity fluctuations without the shear. This was first studied by
\cite{Kra76} in the paradigm of mean-field theory, where he had considered delta correlated
$\alpha$-fluctuations. For strong enough $\alpha$-fluctuations, the net diffusivity can be negative
giving rise to rapid magnetic field growth on all scales, with the small scales growing fastest,
a catastrophe for the mean field dynamo. With finite correlation effects naturally incorporated in
the renovating flow , we revisit this problem.

The response tensor for the mean magnetic field is obtained for the following velocity field (\Eq{vel} without shear)
\beq
\bfv(\bfx, t) = \bfa \sin(\bfq\;\cendot\bfx+\Psi)+h\,\bfc\cos(\bfq\;\cendot\bfx+\Psi)
\eeq
where $\bfq\;\cendot\bfa=\bfq\;\cendot\bfc=0$ and the relative helicity parameter $h=[-1,1]$. Note that this is the same velocity field as considered in \cite{GB92}.
The evolution of the mean-magnetic field in Fourier space is given by \Eq{MatrixEq},
\beq
\widetilde{H}_i(\bfk,n\tau) = {\cal G}_{ij}(\bfk,\tau)\widetilde{H}_{j}(\bfk,(n-1)\tau),
\eeq 
where 

\begin{align}
{\cal G}_{ij}(\bfk,\tau)& = \delta_{ij}g_0(ka\tau,h) + i \frac{hqa\tau}{2k}\epsilon_{ijk} k_k g_1(ka\tau,h) \nonumber \\[1em]
g_0(s,h)& = \overline{j_0(s\chi)}; \quad g_1(s,h)=\overline{\frac{j_1(s\chi)}{\chi}}\\
\chi & = \sqrt{\cos^2\psi+h^2\sin^2\psi}
\end{align}
Here $j_0$ and $j_1$ are spherical Bessel functions of order zero and one, respectively, and $\psi$ is the angle between $\bfk_{\perp}$ and $\bfa$ \citep[see,][for details]{KSS12}. 
The eigenvalues of the response tensor ${\cal G}_{ij}$ are 
\beq
\lambda_0 = g_0; \quad \lambda_\pm = g_0 \pm \frac{h\,q a\tau}{2}g_1,
\label{GBeigen}
\eeq
and the corresponding eigenvectors are 
\beq
(k_1,k_2,k_3); \quad (-ik_2^2-ik_3^2,\; ik_1k_2\mp k k_3,\;ik_2k_3\mp kk_2).
\label{eigvec}
\eeq
We can express the response tensor $\hat{\cal G}$ in the diagonal representation as, 
\beq
\hat{\cal G} = PDP^{-1}
\label{diag}
\eeq
where $D$ is a diagonal matrix: $D={\rm diag}(\lambda_0,\lambda_+,\lambda_-)$ and $P$ has eigenvectors of $\hat{\cal G}$ (see \Eq{eigvec}) in its columns.

We can use the diagonal form representation given by \Eq{diag} to study the helicity fluctuations effect on growth rate, whose renovation time ($\tau_h$) is $m$ times the velocity renovation time: $\tau_h = m\tau$.  The double-averaged response tensor $
\hat{G}$ given by \Eq{tensorG} can be written as
\begin{align}
\hat{G} = &\left\langle {\cal \hat{G}}^m\right\rangle_h = \left\langle (PDP^{-1})^m\right\rangle_h\nonumber \\
 = &\left\langle PDP^{-1}PDP^{-1}\dots PDP^{-1}\right\rangle_h\nonumber \\ 
 = &P \left\langle D^m\right\rangle_h P^{-1} = P\; {\rm diag}(\langle\lambda^m_0\rangle_h\,,\langle\lambda^m_+\rangle_h\,,\langle\lambda^m_-\rangle_h) P^{-1}\,.
\end{align}
The ensuing last simplification occurred because eigenvectors are just the functions of wave vector components $\bfk$.  

\begin{figure*}
\includegraphics[scale=0.4]{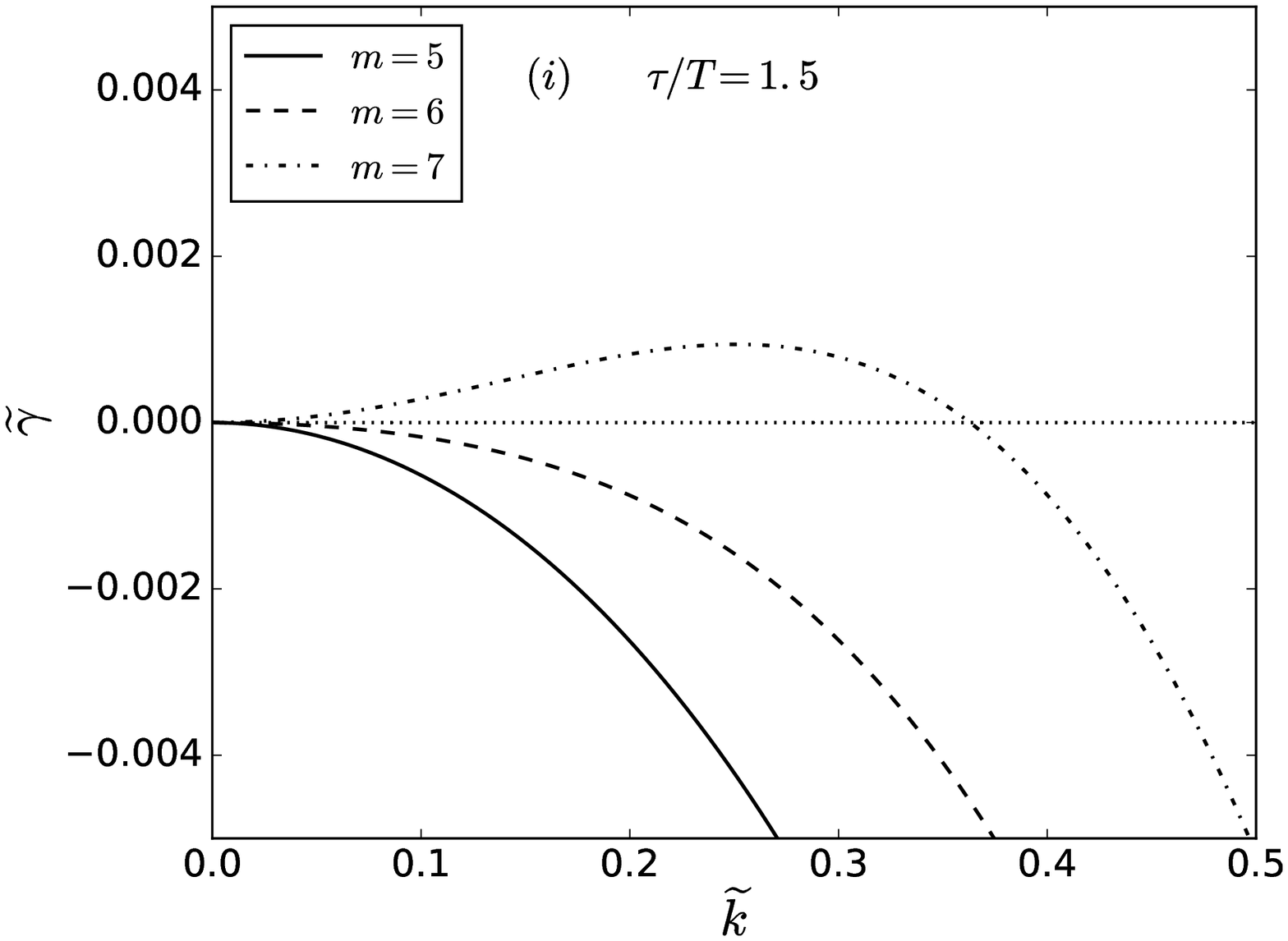}
\includegraphics[scale=0.4]{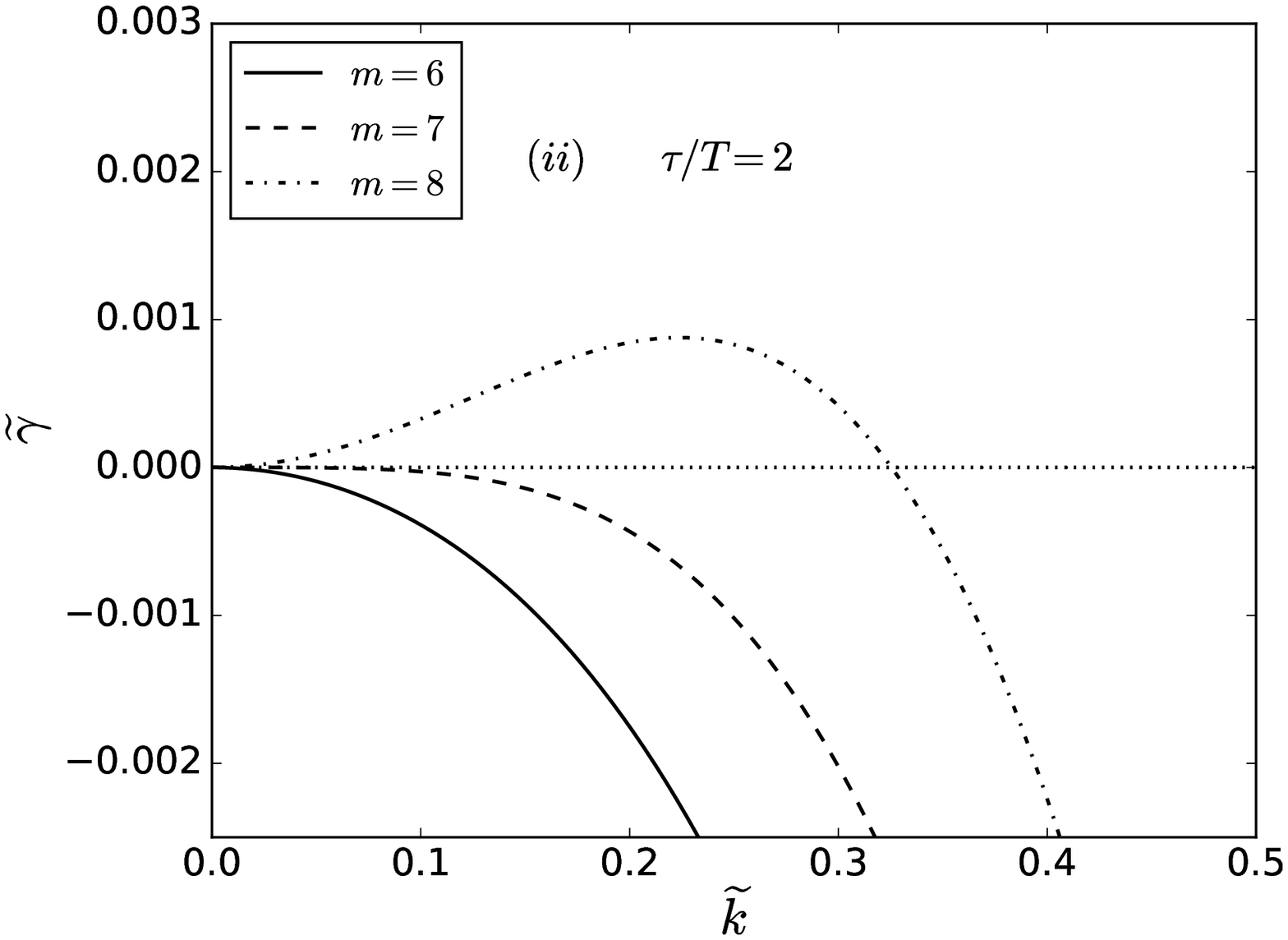}
\caption{Normalised growth rates versus wavenumber for (i) spiked, and (ii) uniform
helicity distributions, where renovation times $\widetilde{\tau}=\tau/T$
are 1.5 and 2, respectively.}
\label{dispersion_plot}
\end{figure*}

If we consider the symmetric distribution of helicity, in which case, the mean and all the odd
moments of helicity are zero, then for such a distribution we have $\left\langle \lambda_+^m \right\rangle_h = \left\langle \lambda_-^m \right\rangle_h$. Using this, the magnetic field at
$n\tau$ can be related to the magnetic field at $(n-m)\tau$ as,
\beq
\widetilde{\bfH}(\bfk, n\tau) = \left\langle \lambda_+^m \right\rangle_h  \widetilde{\bfH}(\bfk, (n-m)\tau)\,.
\eeq
In deriving the above result, we have used the solenoidality condition for the magnetic field ($\bfk\cendot\widetilde{\bfH}=0$).
The growth rate of the mean field is given by (see \Eq{gam_cyc})
\beq
\gamma = \frac{1}{m\tau}\ln \left\langle \lambda_+^m \right\rangle_h\,.
\label{GB_growthrate}
\eeq
Now, we consider two specific distribution for helicity-fluctuations: (i) a \emph{spiked distribution}
with zero value in the range $(-1,1)$ and non zero value at $1$ and $-1$, i.e., PDF$(h) = \frac{1}{2}([\delta(h+1)+\delta(h-1)]$; and (ii) a \emph{uniform distribution} with constant
value $1/2$ in the range $[-1,1]$. Here onwards, we refer to (i) and (ii) simply as spiked and uniform distribution of helicity, respectively.   
We use non-dimensional growth rate $\widetilde{\gamma}=\gamma\, T$ where $T=1/qa$ is a turn over time of the eddy and normalised wavenumber $\widetilde{k}=k/q$ in the figures. 

In \Fig{dispersion_plot}, we have shown the normalized growth rate $\widetilde{\gamma}$ as a
function of normalized wavenumber $\widetilde{k}$ for three different values of helicity
renovation time for each of the distributions. For (i) spiked and (ii) uniform helicity distributions,
we have chosen the value of turn over time to be $\tau/T=1.5$ and $\tau/T=2$, respectively.
The net negative diffusion occurs at $m=7$ for spiked distribution and $m=8$ for uniform distribution,
resulting in the growth of the mean magnetic field in the absence of shear. The largest wavenumbers
(smallest scales) have negative growth rates in the finite time correlated helicity fluctuations in
both the distribution, thus qualifying for a bonafide large-scale dynamo \citep[see][]{JNS18}\footnote{Here finite correlation effects of $\alpha$-fluctuations are
incorporated in the double-averaged mean-field equation, and it shows a similar trend as in
\cite{Kra76} when correlation time of $\alpha$ is finite but small in the absence of shear.}, unlike
as in the white-noise $\alpha$-fluctuations, where largest allowed wavenumbers (smallest allowed
length scales) grow fastest as the growth rate increases monotonically with the wavenumber $k$ (see \Eq{kraich_disper}). Next we will obtain the threshold values for $m$ which lead to negative
diffusion.

\begin{figure*}
\includegraphics[scale=0.45]{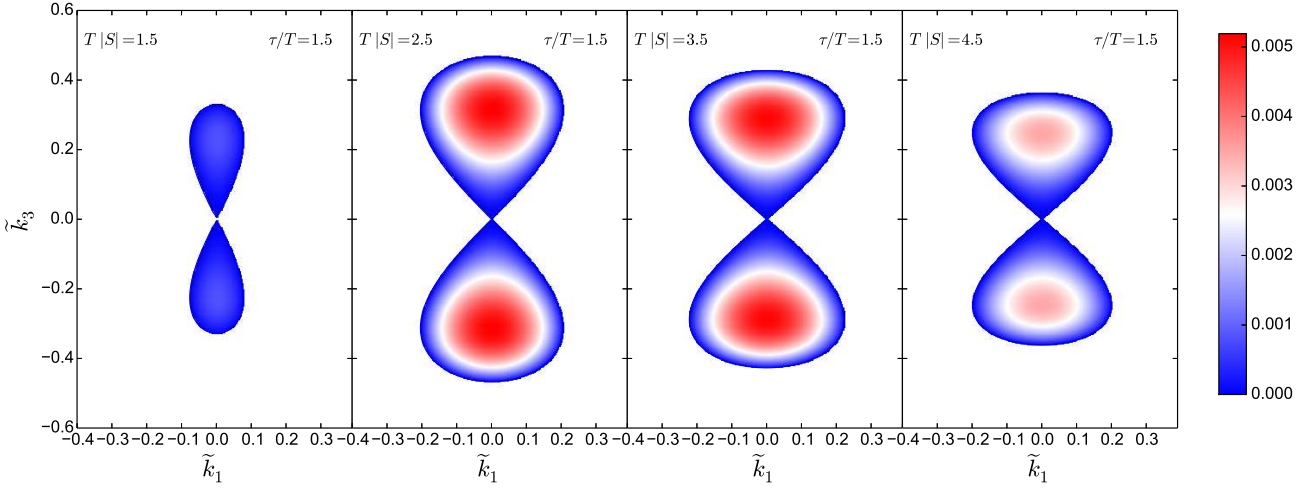}
\caption{Contour plots of the normalized growth rate $\widetilde{\gamma}$ for
axisymmetric mode ($k_2=0$) in the $k_1$-$k_3$ plane, in case of spiked helicity
distribution with $m=3$; shear strength increases from left to right.}
\label{fluchel}
\end{figure*}

\subsection{Comparison with \citet{Kra76}}
\label{kra-sec}
Let us derive the expressions for $\alpha$ and $\eta_t$ from \Eq{GBeigen} in the limit of small
renovation times, such that $ka\tau\ll 1$, which may be though of as a
long wavelength limit at fixed $\tau$ \citep{KSS12}. Expanding \Eq{GBeigen} upto second
order in the dimensionless quantity $ka\tau$ we have, 
\beq
\lambda_\pm = 1-\frac{(ka\tau)^2 \left\langle\chi^2\right\rangle}{6}\mp\frac{(ka\tau)(q a\tau)h}{6}
\eeq
The growth rate is given by $\gamma_\pm=(1/\tau)\ln\lambda_\pm$.
It is convenient to express the transport coefficients in terms of a quantity
\beq
\eta_{}=\frac{a^2 \tau}{3},
\label{etat0}
\eeq
which has the dimensions of the diffusion coefficient.
In the limit of $ka\tau\ll 1$ we obtain
\begin{align}
\gamma_\pm & = -\eta_t k^2 \pm \alpha\,k \\
\mbox{where}\quad
\alpha & = -\frac{\eta_{0} q \tau h}{2}; \quad \eta_t = \frac{\eta_{0}}{4}(1+h^2)\,. 
\label{alpha_and_eta}
\end{align}
To make a comparison with \cite{Kra76} who essentially assumed delta--correlated--in--time
$\alpha-$fluctuations\footnote{Delta--correlated--in--time or the so-called \emph{white--noise}
regime may also be thought of as a limit in which $\eta_{t0}$ approaches a fixed nonzero
value as $\tau\to 0$.} \citep{SS14}, we first expand the eigenvalue $\left\langle \lambda_+^m \right\rangle_h$ for double-averaged mean magnetic field upto $O(ka\tau)^2$:
\begin{align}
\left\langle \lambda_+^m \right\rangle_h\; = &\; 1 -\frac{m}{6}(k a \tau)^2\left\langle\chi^2\right\rangle_h + \frac{m(m-1)}{72}(ka\tau)^2\,(qa\tau)^2\,\left\langle h^2\right\rangle_h
\end{align}
Then the growth rate $\gamma$ as given by \Eq{GB_growthrate} can be written explicitly
in the limit $ka\tau \ll 1$ as

\begin{align}
\gamma & = -\left(\left\langle\eta_t\right\rangle_h - \left\langle\alpha^2\right\rangle_h\frac{(m-1)\tau}{2}\right)k^2 \label{kraich_disper}\\[1em]
& \mbox{where} \quad \left\langle\eta_t\right\rangle_h = \frac{\eta_{0}}{4}(1+\left\langle h^2\right\rangle_h) \quad \nonumber \\[1em]
& \mbox{and}\quad \left\langle\alpha^2\right\rangle_h = \left(\frac{ \eta_{0}q}{2}\right)^2\left\langle h^2\right\rangle_h\,,
\label{kraich_alpha_fluc}
\end{align}
where $\eta_{0}$ is defined in \Eq{etat0}.
Since helicity renovates after every $m\tau$, the transport coefficients $\alpha$
and $\eta_t$, as given in \Eq{alpha_and_eta}, also fluctuate in time. Therefore we have
defined the average of them over the helicity fluctuations.  
The dispersion relation in \Eq{kraich_disper} is similar
to the one obtained in \cite{Kra76} for delta correlated $\alpha$-fluctuations. The second term in the dispersion relation is non-zero, only when $m>1$, i.e., when helicity correlation time exceeds
the velocity correlation time. This causes reduction in the net turbulent diffusion, and
mean magnetic fields can grow exponentially if net diffusion becomes negative. The strength
of the second term is directly proportional to the strength of the helicity fluctuations
as well as the correlation time of the helicity fluctuations, as may be seen from
\Eq{kraich_alpha_fluc}. 

From \Eq{kraich_disper}, we can determine the condition on $m$, so that it would lead
to net negative diffusion as:
\beq
m-1 > \frac{12\left\langle \chi^2\right\rangle_h}{(qa\tau)^2\left\langle h^2 \right\rangle_h}\,.
\eeq
For $h=\pm 1$ (spiked distribution) and $qa\tau = 1.5$, we have
$\left\langle\chi^2\right\rangle_h =1$, yielding $m>6$, i.e., if the helicity renovation
time is more than six times the velocity renovation time, then it leads to the negative diffusion,
triggering the mean-field dynamo instability. 

Similarly, for uniform distribution of helicity in the range $[-1,1]$ and $qa\tau = 2$, we have $\left\langle\chi^2\right\rangle_h =2/3$ and $\left\langle h^2\right\rangle_h =1/3$, in which
case, $m>7$ for negative diffusion to occur; \Fig{dispersion_plot}. In the next section, we will explore the case with shear where we focus on the regime when the value of $m$ is below its threshold value for negative diffusion, by considering $m<6$ and $m<7$ for spiked and uniform distributions of helicity fluctuations, respectively.

\section{helicity fluctuation with shear}
\label{HFWS}
All the important relations to study the dynamo action for large-scale magnetic fields
due to zero-mean helicity fluctuations in presence of shear is derived in \Sec{EMF}.
The response tensor, as given in \Eq{tensorG}, governs the evolution of the
mean magnetic field, and its eigenvalues provide the growth rate and cycle period of the
dynamo (see \Eq{gam_cyc}).
To determine the double-averaged response tensor $\hat{G}$, given in \Eq{tensorG}, we have
to first obtain the response tensor $\hat{\cal G}$ of \Eq{res-tens} which includes
contributions from the shearing plane waves (\Eq{vel}). The averaging over the phase
$\Psi$ of the wave could be performed analytically \citep[see][for details]{KSS12,JS20},
providing,
\beq
{\cal G}_{ij}(\bfK,\tau)=\gamma_{il}(\tau)\left\langle \delta_{lj} J_0\left( \Delta\right)-ih
\frac{q_j\left[\bfK\cross(\widetilde{\bfa}\cross\widetilde{\bfc})\right]_l }{\Delta}
J_1\left( \Delta\right)\right\rangle_{\bfq,\bfa}
\label{shear_turbulence_tensor}
\eeq
where
\begin{align}
\widetilde{\bfa}(t,\bfq)& = \displaystyle\int\limits_{(n-1)\tau}^{n\tau}  [\bfA - S (t-t_0) A_1\ey]\; \ud t\,,\\
\widetilde{\bfc}(t,\bfq) & = \int\limits_{(n-1)\tau}^{n\tau}  [\bfC - S (t-t_0) C_1\ey]\; \ud t\,,
\end{align}
and $\Delta= \sqrt{(\bfK\cdot\widetilde{\bfa})^2+h^2(\bfK\cdot\widetilde{\bfc})^2}$;
$J_0$ and $J_1$ are Bessel functions of order zero and one, respectively. Here $\bfK$ is the magnetic field wavevector at time $(n-1)\tau$, i.e., $\bfK_{n-1}=[k_1-S\,\tau(n-1)k_2,k_2,k_3]$. The averaging over the direction of the initial wavevector $\bfq$ and initial amplitudes $\bfa$ and $\bfc$ is performed numerically; the algorithm to perform the averages is given in the
Appendix~\ref{Aveg_tech}. The term that is relevant for the dynamo action is the second term in \Eq{shear_turbulence_tensor}, because $J_0(y)\leq 1$, $\forall\;\; y$. If $h=0$, i.e.,
strictly non-helical case when the helicity vanishes point-wise, then there is no mean-field
dynamo. Even if we allow for the symmetric helicity fluctuations at this point, second term vanishes because it is odd in $h$.
\emph{Therefore, even in the presence of shear,
for $m=1$ case, i.e., when the fluctuation time-scales of helicity and velocity are the same,
there is no mean-field dynamo action, a conclusion which is similar to the case when shear is absent; see also \Eq{kraich_disper}.}
Hence we obtain the double-averaged response tensor $\hat{G}$ given in \Eq{tensorG} for the case $m>1$.

\begin{figure}
\includegraphics[scale=0.4]{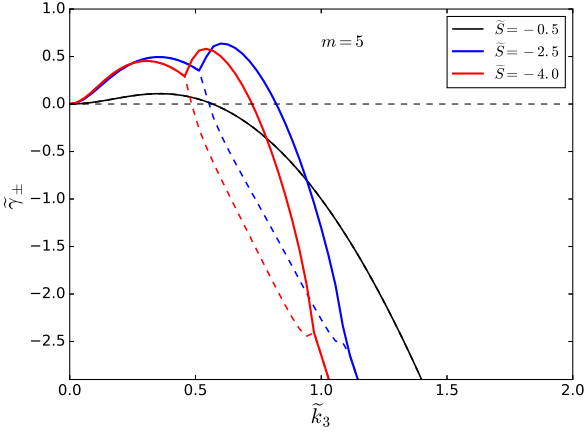}
\caption{Growth rates corresponding to the two eigenvalues as functions of the
vertical wavenumber $k_3$, for three choices of shear rate $S$.
Bold (dashed) curves represent $\widetilde{\gamma}_+$ ($\widetilde{\gamma}_-$).
Here we have considered the uniform helicity distribution in the range $[-1,1]$.}
\label{lampm}
\end{figure}

As we have seen in the  \textsection~3 of paper~I, if the eigenvalue of the response tensor depends on the particular interval, i.e., if it depends on $n$, then we had concluded that the non-axisymmetric mode of the magnetic field will decay eventually. Physically, it means that the wavevector $\bfK_n$ will increase in time as the magnetic field is evolved across the intervals, creating thus
the smaller scale magnetic structures which ultimately decay by
turbulent diffusion. The situation is similar in this case too with fluctuating
helicity, and therefore, from now onwards we set $k_2=0$, and investigate
the behaviour of only the axisymmetric dynamo.

\begin{figure*}
\includegraphics[scale=0.4]{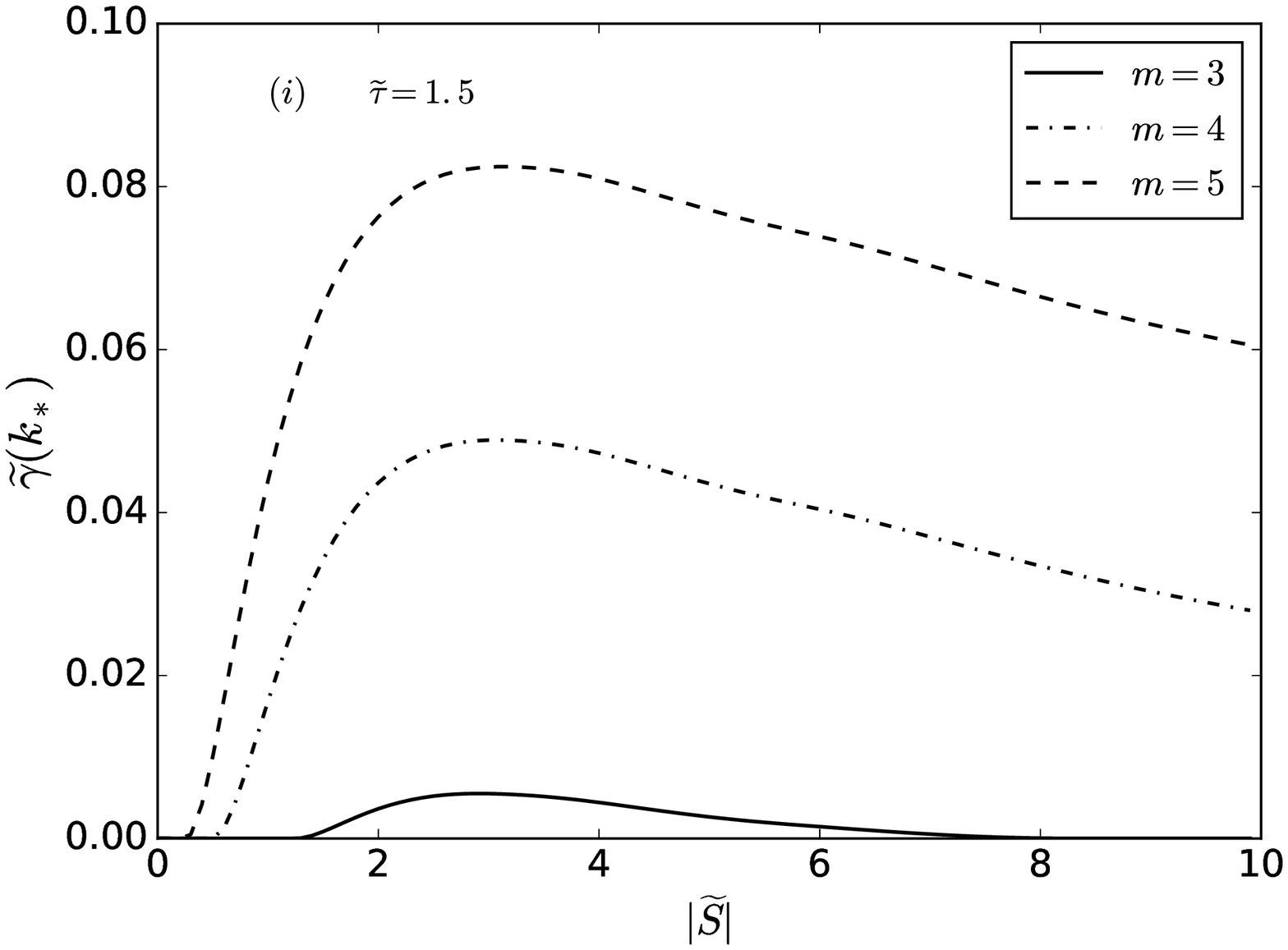}
\includegraphics[scale=0.4]{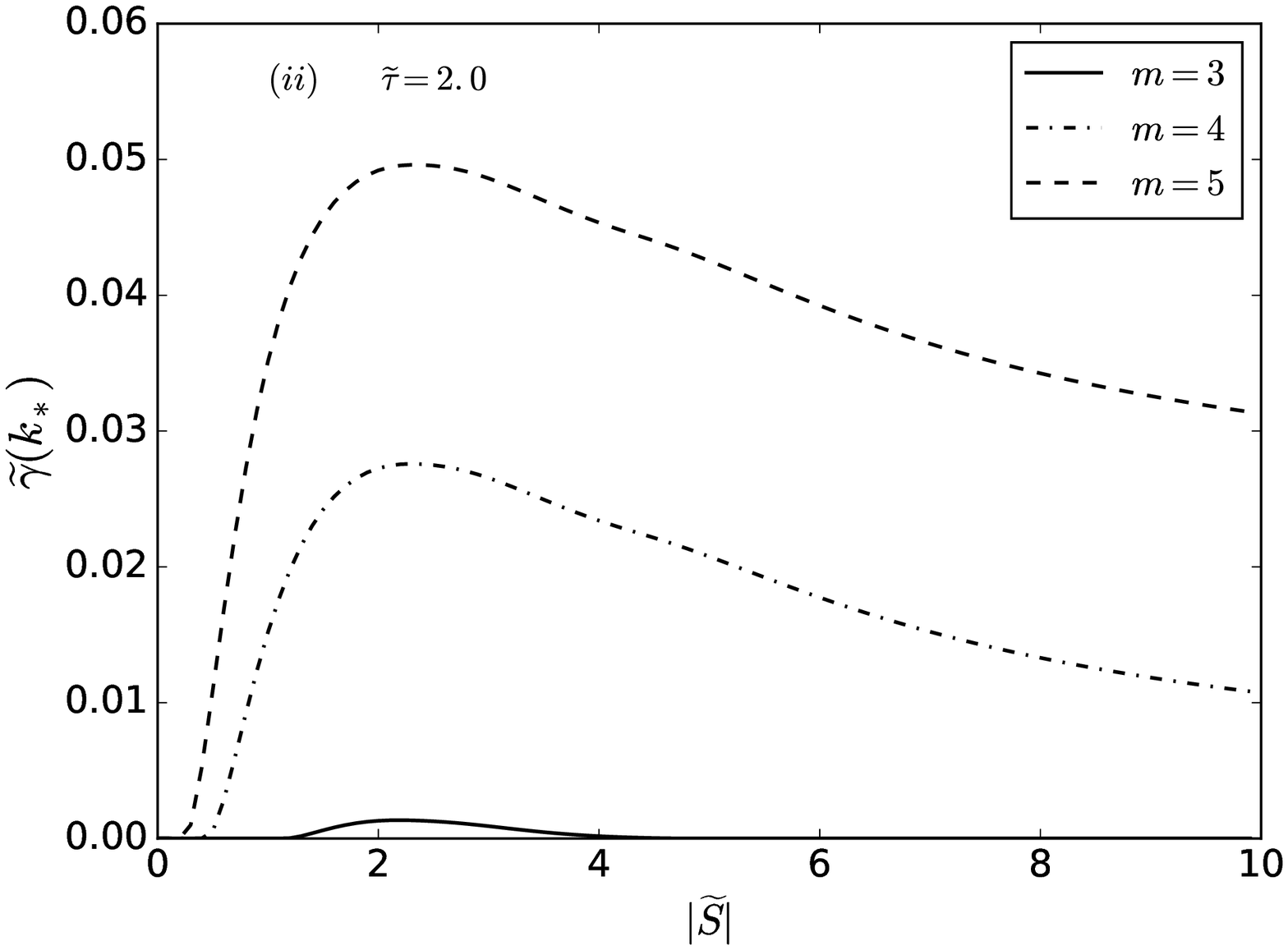}
\caption{Shear dependence of normalized growth rate $\widetilde{\gamma}(k_\ast)$
for spiked (left) and uniform (right) helicity distributions, with $\widetilde{\tau}$
being 1.5 and 2, respectively.}
\label{fluc_growth_rate}
\end{figure*}

In \Fig{fluchel}, we show the normalized growth rate ($\widetilde{\gamma}$)
in the $k_1-k_3$ plane with $k_2=0$. Shear strength is increased from left to
right while keeping the turnover time as fixed $(\tau/T =1.5)$.
The positive growth rate region is referred to as a dynamo region -- a region enclosed within the outermost blue contours. The plots are for spiked helicity distribution $(h=\pm 1)$ with $m=3$, as we have verified that this is the
minimum value of $m$ that is needed to trigger the dynamo instability, and it
is well below the threshold ($m=6$) for the negative diffusion. \emph{Thus,
we find the possibility of mean-field dynamo action without the need for
negative diffusion of the magnetic field.} The dynamo region first expands
as shear increases but then it begins to shrink as shear is enhanced further;
this nature revealing the dynamo quenching at very strong shear becomes
more apparent in \Fig{fluc_growth_rate}.
As may be seen from the contour plots of \Fig{fluchel}, the maximum growth
occurs on the line $k_1=0$, and the dynamo region is symmetric about $k_3=0$.
Therefore, we set $k_1=0$ as well, without any loss of generality, and
henceforth study the one-dimensional modes of dynamo along the positive values
of $k_3$; this is equivalent to taking the $X_1-X_2$ spatial average, normally
used in numerical works. 

\begin{figure}
\includegraphics[scale=0.4]{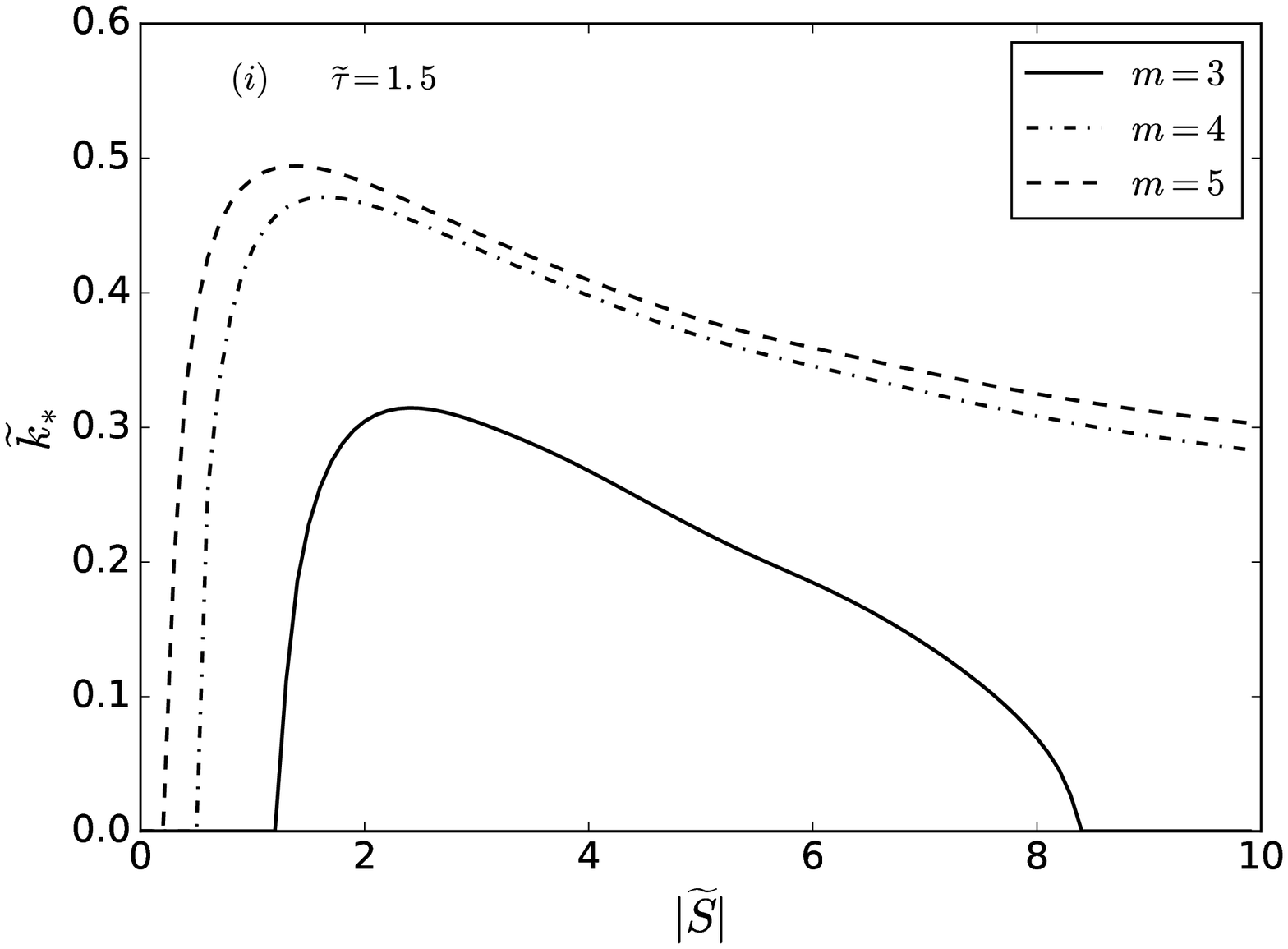}
\includegraphics[scale=0.4]{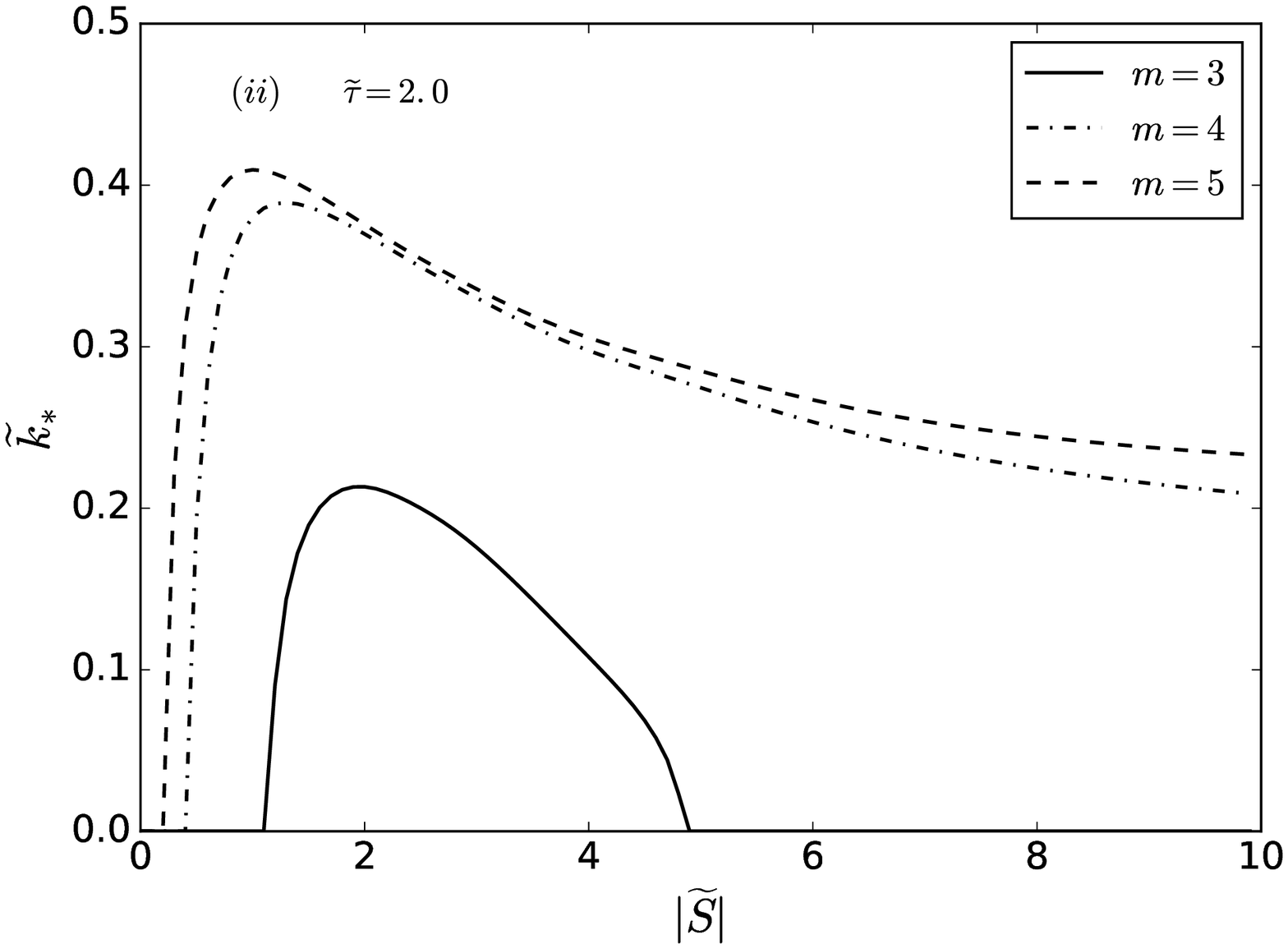}
\caption{Shear dependence of maximally growing wavenumber $\widetilde{k_\ast}$
for spiked (top) and uniform (bottom) helicity distributions, with $\widetilde{\tau}$
being 1.5 and 2, respectively.}
\label{fluc_k3max}
\end{figure}

In \Fig{lampm}, we show the growth rates obtained from both the eigenvalues of the double-averaged response tensor, \Eq{tensorG}, for the case of
uniform helicity distribution with $m=5$. We have chosen three different
values of shear strengths. At $\widetilde{S}=-0.5$, both the eigenvalues
are same for symmetric helicity distribution, but, as shear increases we
see a bifurcation of the eigenvalues at certain wavenumber $k_3$, and
the corresponding eigenvalues overlap again when $k_3$ is sufficiently
large; see, e.g., dashed and solid red (or blue) curves in \Fig{lampm}.
This bifurcation in eigenvalues is likely due to the artificial nature of the
renovating flows, where we had assumed that the velocity field across the successive intervals are independent of each other, which destroys the continuous time-translational symmetry and makes it discrete over the times $n\tau$ with $n=1,2,3\dots$. Therefore, to compute the maximum growth rates and the corresponding maximally growing wavenumber, denoted by $k_\ast$, we focus on
the peak which is before the bifurcation, where both the eigenvalues have same value.

\begin{figure*}
\includegraphics[scale=0.4]{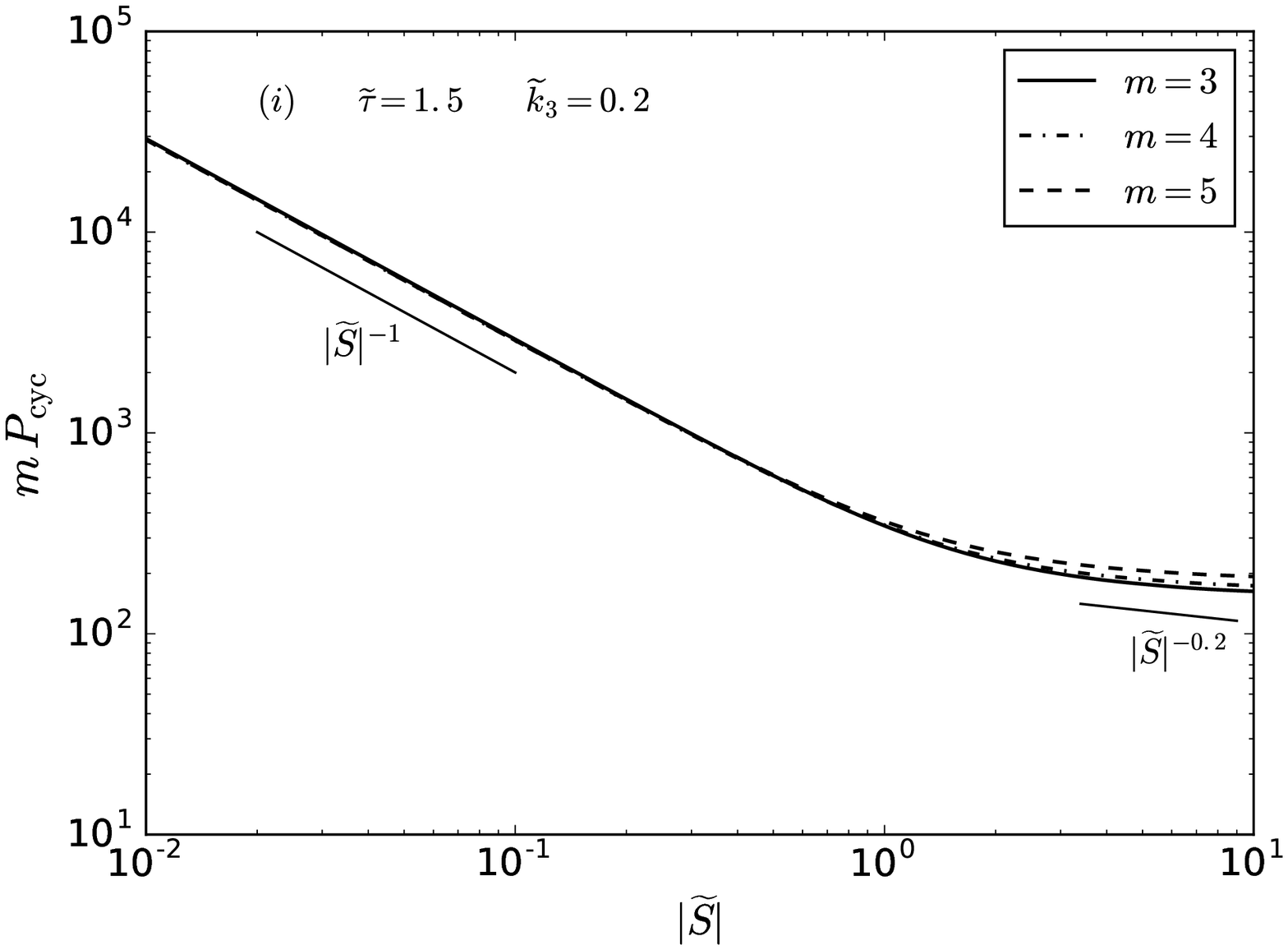}
\includegraphics[scale=0.4]{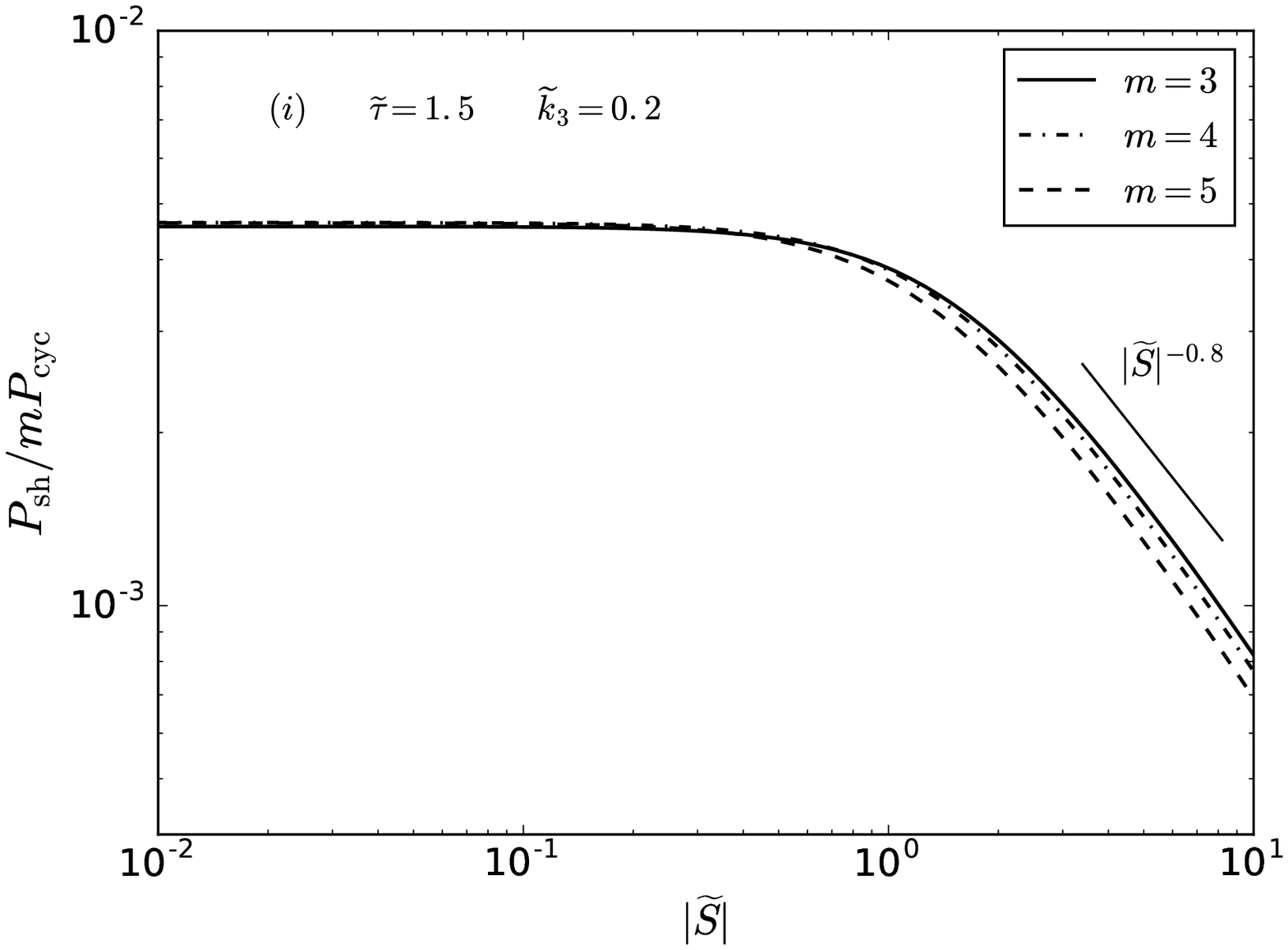}
\caption{Shear dependence of scaled cycle period of the dynamo wave (left), and
normalised ratio $P_{\rm sh}/m P_{\rm cyc}$ (right), at fixed $k_3=0.2$ and
$\widetilde{\tau}=1.5$; $P_{\rm sh}=1/|S|$.}
\label{pcyc_helfluc}
\end{figure*}

\Fig{fluc_k3max} shows the normalised wavenumber corresponding to the fastest growing mode, denoted by $k_\ast$, as a function of shear.
The maximum wavenumber $k_\ast$ varies non-monotonically with the shear
$|S|$, when the renovation time $\tau$ is comparable to the eddy turnover time $T$. This is similar to the trend observed for fixed helicity case in Paper~I.
The left hand panel of the figure is plotted for (i) spiked helicity distribution with renovation time $\tau = 1.5\,T$ and the right hand panel is for (ii) uniform helicity distribution with $\tau = 2.0\,T$. For both the plots, we have considered three different values of $m$,
all below the threshold for negative diffusion of the mean magnetic field. The mean field grows maximally at scales greater than the eddy scale $1/q$ for whole range of shear value considered. This qualifies for the bonafide large-scale dynamo for symmetric helicity distribution with zero mean without any negative diffusion. 

Now let us look at the dependence of growth rate on the shear. We saw that the wavenumber
corresponding to the fastest growing mode is itself a function of shear. In
\Fig{fluc_growth_rate}, we show the growth rate $\gamma$ at $k_\ast$ as a function of shear. Interestingly, the growth rate shows a non-monotonic behaviour with respect to the shear strength $|S|$, for both the helicity distributions; spiked (uniform) distribution with renovation time
$\tau = 1.5 T$ ($\tau= 2.0 T$). Here, the growth rate increases linearly with shear strength $|S|$, it reaches a maximum before it starts to decrease when shear is too strong. This trend is seen in all three different curves plotted for renovation time of helicity fluctuations in the range $3T\leq m\tau\leq 5T$. Linear variation with shear for moderate shear strengths is observed in
previous numerical studies on the shear dynamo \citep{You08b,SJ15}. In light of the present work, this suggests that the cause of dynamo in those simulation could be helicity fluctuations coupled with large-scale shear. When $0\leq|\widetilde{S}|\leq 3$, slopes of the curves in \Fig{fluc_growth_rate} depend strongly on the correlation time of helicity fluctuations, with
slopes increasing as the value of $m$ is made larger. \cite{You08b} noted that the proportionality constant in the relation $\gamma \propto |S|$ is neither the function of fluid nor the magnetic Reynolds numbers. This would be consistent with our findings in this paper where the slope is
quite sensitive to the helicity correlation time.

Cycle period of the dynamo wave, as obtained from \Eq{gam_cyc}, is shown in
\Fig{pcyc_helfluc}, where, for illustrative purposes, we just consider the case of
spiked helicity distribution. 
Interestingly enough, if we plot $m\,P_{\rm cyc}$ for a fixed vertical wavenumber ($\widetilde{k}_3=0.2$) versus the shear strength $|S|$ (left panel of \Fig{pcyc_helfluc}), different curves corresponding to the different values of $m$ collapse on top of each other.
At the smaller value of shear, $m\,P_{\rm cyc}\sim |S|^{-1}$ and as shear strength increases the slope decreases to $|S|^{-0.2}$. This becomes even more apparent as we plot the ratio $P_{\rm sh}/m\,P_{\rm cyc}$, where $P_{\rm sh}=1/|S|$, as a function of shear strength in the right hand panel of \Fig{pcyc_helfluc}. Thus our model yields, for a dimensionless quantity $1/m\,P_{\rm cyc}|S|$, a scaling behaviour of (a) $|S|^{0}$, that is independent of shear, at weak shear,
and (b) $|S|^{-0.8}$ for stronger values of shear.

\section{Discussion}
\label{Diff}

\begin{figure}
\hspace{-0.5cm}
\includegraphics[scale=0.4]{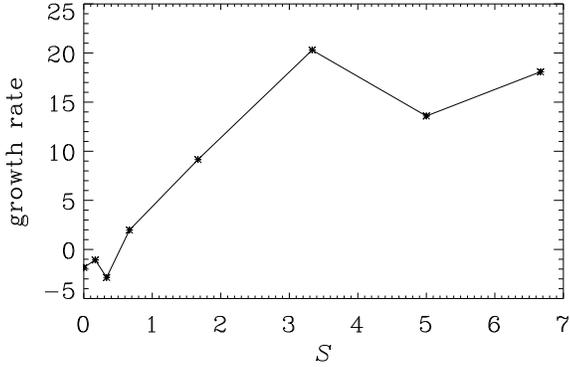}
\caption{Plot of large-scale magnetic field growth rate obtained at box scale versus shearing rate from \citet{DP09}\copyright APS. Reprinted with permission.}
\label{dp09}
\end{figure}

Figure~\ref{dp09}, taken from \cite{DP09}, shows the shear dependence of dynamo
growth rate, calculated at the scale of the box, from a setup of stratified convective slab with rotation vector
pointing vertically and shear in the horizontal plane.
The $\alpha$-effect in their system is sub-critical to trigger the dynamo, as
may also be seen by the negative growth rate for zero shear in Figure~\ref{dp09}.
Although their system is different from ours, it is shown \citep{CH06,TCN08,DP09} that shear
could be more important component to generate the LSD than the combined
effect of stratification and rotation.

\begin{figure}
\includegraphics[scale=0.4]{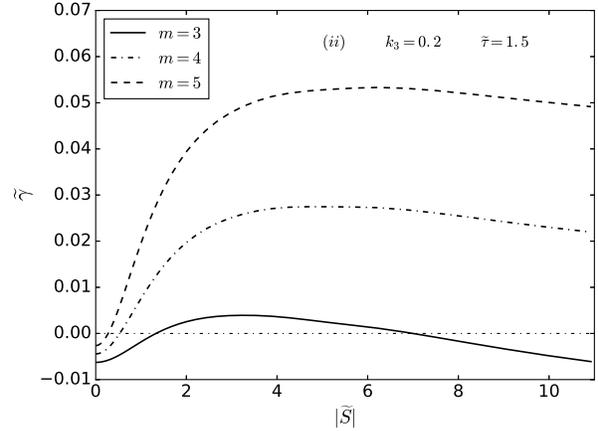}
\caption{Plot of growth rate as a function of shear strength for fixed value of wavenumber.}
\label{fixed_k}
\end{figure}

In Figure~\ref{fixed_k}, we show the normalised growth
rate as a function of shear $|S|$ for a fixed wavenumber. Note that
the growth rates are scaled differently in simulation. Nevertheless, we observe a
similar qualitative trend in both these figures: growth rates are negative at zero
shear, indicating that there is no negative diffusion; positive growth rates are obtained
when shear becomes stronger than some threshold shear rate, beyond which these
increase linearly with shear upto some shear rate, before showing saturation
for higher value of shear. In our case, the threshold is a function of helicity
correlation time ($m\tau$) and $m=4$ case matches with the threshold in Figure~\ref{dp09}.
At very strong shear, we observe dynamo quenching in Figure~\ref{fixed_k}, which is likely due to
the quasi-two-dimensionalization of the stochastic flow, as the $u_1$ component is
expected to be rapidly sheared out when shear becomes too large; see Figure~\ref{vel},
and Appendix~\ref{anti_dynamo} for the relevant anti-dynamo theorem.

Most studies on the mean-field dynamos employ either ensemble or spatial/temporal averages.
\citet{Hoy88} showed that the spatial or temporal averages obey Reynolds rules only
approximately where the error is proportional to the correlation time of the turbulence.
This error appears as a forcing term in the mean-field equation causing fluctuations in
the transport coefficients.
In solar context, \cite{Hoy93} argued that the fluctuation in $\alpha$
is of the order $\sim u_0/\sqrt{N}$, where $u_0$ is the turbulent velocity and $N$ is the
number of convective cells over which averaging is performed.

Fluctuations in transport coefficients could have more fundamental origin, regardless of the
averaging scheme. \citet{Kra76}, for example, reported the possibility of a reduction in the
turbulent diffusion due to the presence of $\alpha$ fluctuations, which in this case originate
because of fluctuations in the kinetic helicity of the turbulent flow; see \Eq{kraich_disper} and
note that sufficiently strong $\alpha$ fluctuations can, in principle, trigger a dynamo by
making the turbulent diffusion as negative.
The kind of fluctuations in the transport coefficient as discussed in \cite{Hoy88} are due
to the non-equivalence of the spatial/temporal averages with ensemble averaged fields.
However, in the present work we have shown the growth of the ensemble averaged mean-field,
therefore, such fluctuations are, by definition, absent in our work.

In this work, the helicity fluctuations are the source of fluctuations in the transport
coefficients. Starting from the induction equation, we obtained the Kraichnan's equation for
mean-field in the case of zero shear and make comparisons in \Sec{kra-sec}.
\cite{Kra76} considered a special type of turbulence which he called as \emph{non-normal turbulence}
where helicity fluctuates over times that are longer than the velocity correlation times.
In any turbulent system, derived quantities, such as the kinetic helicity, are also expected
to show fluctuations, possibly over times different from those of the velocity fields.
As the horizontal averages are routinely employed in numerical works \citep[e.g.][]{BRRK08,AbSu19},
both types of fluctuations described above are expected to be present and they could contribute
to the measurements of the transport coefficients and their fluctuations. It may be useful
to separate the contributions arising from the more fundamental origin in terms of helicity
fluctuations. Moreover, it is important to quantify the time scale separation between velocity and
helicity correlation times in the future numerical works, as this non-normal turbulence discussed in \cite{Kra76}
could indeed play a critical role in the operation of the shear dynamo. 

\section{Conclusions}
\label{Conclude}
We have studied the effects of zero-mean helicity fluctuations that are distributed
symmetrically about zero on the growth of large-scale magnetic field in presence of
a background linear shear flow. To study this, we have extended the renovating flow
of \citep{GB92} to include the shear as well as the helicity fluctuations whose renovation
time ($\tau_h$) is different from that of the velocity renovation time ($\tau$).
Single helical waves, that are also the exact solutions to the Navier-Stokes equation,
were chosen in our model. Anisotropic effects of shear are thus self-consistently included
in the random renewing flow, and as a result, also on the transport properties.
One of the amplitudes of the velocity, the $A_1$ component, that is in the direction of
linear variation of the shear (i.e., perpendicular to the direction of the shear),
is continuously sheared to produce the other two components, thus making the flow quasi-two
dimensional at strong shear, which has serious implication for dynamo action due to the
anti-dynamo theorem discussed in Appendix~A.

For the velocity field, we consider the same ensemble which was used in the paper~I.
Following \citet{Kra76,Sok97}, we employ here a double-averaging scheme to determine the
evolution of mean magnetic field: (i) first the averaging is performed over the parameters
of the velocity which take random values in different renovating intervals of length
$\tau$, and (ii) then over the realizations of the kinetic helicity which is allowed to
vary over the interval of $m\tau$, where $m\geq 2$. Thus, the response tensor, also
called the Green's function, which maps the magnetic field from $(n-m)\tau$ to $n\tau$, is
obtained. Its eigenvalues determine the dispersion relation, from which the growth rate
($\gamma$) and the cycle period ($P_{\rm cyc}$) of the mean-field dynamo wave are determined. 

For the case of zero shear, we obtained the threshold values of $m$ (i.e., $\tau_h/\tau$),
such that, the larger values make the turbulent diffusion negative, leading to the growth of
mean magnetic field \footnote{This is same as the strong $\alpha-$fluctuations discussed in
\citet{SS14,JNS18}.}.
By expanding the response tensor to $O(ka\tau)^2$ in the limit of small correlation time,
i.e., when $ka\tau\ll 1$, we derived the dispersion relation and obtained explicit expressions for
the transport coefficients. These results are in agreement with \cite{Kra76}, where
dynamo is possible only by the process of negative turbulent diffusion, i.e., when
helicity fluctuations are sufficiently strong.

We then studied more general case with finite shear and $m\geq2$, i.e., when
the helicity correlation times are longer than that of the velocity field, which
is taken to be on the order of the eddy turnover time $T$.
This work thus further generalizes the work of \cite{JNS18} to include the non-local
effects with memory and, in a sense, role of tensorial $\alpha-$fluctuations, by
self-consistently incorporating the effect of shear on the stochastic velocity and helicity
fields.

Non-axisymmetric modes ($k_2\neq 0$) of mean-field dynamo decay eventually as $t\to \infty$,
just like in the case of fixed helicity studied in paper~I.
Therefore, we focus on the axisymmetric ($k_2=0$) mean-field dynamo and list our notable
findings in the following:

\begin{itemize}
\item [(a)] The challenge was to show the growth of magnetic field in shearing flows with
helicity fluctuations that are distributed symmetrically about zero, \emph{without any negative
turbulent diffusion} of mean magnetic field. We found the growing dynamo solutions
for $m$ which were well below the threshold for negative diffusion;
the threshold was determined in \Sec{HF} for the case of zero shear.
By choosing two types of zero-mean helicity distributions, called spiked and uniform,
we found, in both cases, that the minimum time scale separation needed to trigger the dynamo is
$m=\tau_h/\tau=3$. Minimum value of the shear rate $S$ needed for the dynamo action decreases
as we increase $m$; see \Fig{fluc_growth_rate}. 
\\
\item [(b)] Growth rate corresponding to the fastest growing mode, $\gamma(k_\ast)$, and
the fastest growing wavenumber, $k_\ast$, show non-monotonic behaviour with $|S|$,
for both types of helicity distributions, when velocity renovation times are on the
order of the eddy turnover times. At small shear, the growth rate shows a linear trend, i.e.,
$\gamma \sim |S|$, but in strong shear regime, it starts to decrease as $|S|^{-0.4}$.
The dynamo is thus quenched for sufficiently strong values of shear. This
may be understood in terms of rapid quasi two-dimensionalization of the flow $\bfu$ within
each renovation time interval, when shear becomes too large, as in this case the $u_1$ component
is expected to approach zero as $|S| \to \infty$; see \Fig{vel_amp} and note that
such flows cannot support a dynamo due to the \emph{anti-dynamo theorem} discussed in
Appendix~\ref{anti_dynamo}.
\\
\item [(c)] The growth rate $\gamma$ first increases with wavenumber $k$, and becomes negative
at sufficiently large wavenumbers after it reaches a maximum. For the whole range of shear
strengths that we explored, the fastest growing scale $(k_\ast^{-1})$ is always found to be
larger than the injection scale ($q^{-1}$) of the fluid kinetic energy,
thus enabling a bonafide large-scale dynamo. 
\\
\item [(d)] Dynamo cycle period $P_{\rm cyc}$ scales with shear rate as
$P_{\rm cyc} \propto |S|^{-1}$ at weak shear, with its dependence becoming shallow at large
$|S|$. It thus shows a qualitatively similar trend as we saw in case of fixed kinetic helicity
\citep{JS20}. Different curves for various choices of $m$ collapse on top of
each other when we plot $m P_{\rm cyc}$ as function of $|S|$, indicating that at fixed
shear, $P_{\rm cyc} \propto 1/m$.
\end{itemize}

Our model, which includes memory effects and
the influence of shear also on the stochastic flow in a self-consistent manner,
is successful in
predicting a linear scaling for the dynamo growth rate with shear at small values
of shear rates,
purely by temporal fluctuations in the kinetic helicity that are weak enough to
avoid any negative turbulent diffusion. This scaling relation is in agreement
with earlier numerical works which are able to explore only moderate shear strengths
\citep[see, e.g.,][]{BRRK08,You08b,SJ15}. Another important prediction that the dynamo
is quenched when shear is too large also appears to be a reasonable expectation.
Key assumption in this model is that $\tau_h > \tau$, i.e., the kinetic helicity varies over
times that are slower as compared to the stochastic velocity field.
It is known that the correlation times in turbulent flows are, in general, scale
dependent, and it is indeed likely that $\tau_h(k) > \tau(k)$, at least for a range
of values of $k$ in the anisotropic turbulence created in background rotating and shearing flows, which may be enough to trigger an LSD as predicted in this work.
Determining these correlations times and their scale dependence to assess the magnitude
of time-scale separations will be quite important to further understand the shear dynamos.

\bibliographystyle{mn2e}
\bibliography{ms}

\onecolumn
\appendix

\section{Anti-dynamo theorem in shearing flows}
\label{anti_dynamo}
As we have seen in this section \textsection~\,\ref{Mod} and also in \Fig{vel_amp}, the first component (component in the direction of $X_1$) of the turbulent velocity field $\bfu$ is sheared out by the mean flow given by $SX_1\ey$. The question we are interested is, can we extend the anti-dynamo theorem of \cite{Zel56} to the velocity field $\bfu = (0,u_2,u_3)$, where $u_2=u_2(x_1,x_2,x_3,t)$, with the background shear flow? 
The induction equation with the background shear flow is given by
\beq
\left(\frac{\partial}{\partial t} + SX_1\frac{\partial }{\partial X_2}\right)\bfB +(\bfu\cendot\bnabla)\bfB-S\,B_1 \ey = (\bfB\cendot\bnabla)\bfu + \eta\nabla^2 \bfB\,.
\label{b11_eq}
\eeq
The equation for the $B_1$ component can be separated out as, 
\beq
\left(\frac{\partial}{\partial t} + SX_1\frac{\partial }{\partial X_2}\right)B_1 +(\bfu\cendot\bnabla)B_1 =  \eta\nabla^2 B_1
\label{b1_eqn}
\eeq
 The \Eq{b1_eqn} resembles the advection-diffusion equation for the scalar. If we assume that the magnetic field vanishes at infinity, we can write the above equation as 
 \beq
\frac{\partial}{\partial t}\int \half B_1^2\ud V + \int \bnabla\cendot[\half(SX_1\ey +\bfu) B_1^2]\ud V = \eta \oint \half \bnabla (B_1)^2\cendot \ud\bfS -\eta \int |\nabla B_1|^2 \ud V\,.
\eeq
If the magnetic field vanishes at infinity or if the velocity field is considered to be in periodic domain \cite[see section 6.8,][]{Mof78}, then the surface integrals on L.H.S and R.H.S vanishes. When $\eta=0$, we will have neither growth or decay of the magnetic field, the initial value remains. When $\eta>0$, then we will have exponential decay of the first component of the magnetic field, i.e., $B_1=0$. If we introduce the vector potential $\bfA$ as $\bfB = \bnabla\cross\bfA$, then we can uncurl the \Eq{b11_eq} to get,  
 
 \beq
 \frac{\partial\bfA}{\partial t} + SX_1B_1\ez = SX_1B_3\ex + (\bfu\cross\bfB) + \eta\nabla^2 \bfA + \bnabla\phi\,.
 \eeq  

 Since, we have gauge freedom to transform $\bfA\to \bfA +\bnabla \chi$, we can choose $\chi$ such that 
 $\partial \chi/\partial t - \eta \nabla^2 \chi =\phi$. Therefore we have

\beq
\frac{\partial\bfA}{\partial t} + SX_1B_1\ez = SX_1B_3\ex +  (\bfu\cross\bfB) +  \eta\nabla^2 \bfA
\eeq
 Now, we will write the above equation in component form by setting $u_1=0$ and $B_1=0$ as,
\begin{align}
& \frac{\partial A_1}{\partial t} =\;SX_1B_3+  u_2B_3-u_3B_2 + \eta\nabla^2 A_1\,, \label{A1} \\[1em]
& \frac{\partial A_2}{\partial t} =\; \eta\nabla^2 A_2\,, \quad \frac{\partial A_3}{\partial t} =\; \eta\nabla^2 A_3  \label{A2}
\end{align}

The $A_2$ and $A_3$ components obey the diffusion equation, therefore they must  go to zero. Since, $\bfB = \bnabla\cross\bfA$, with $A_2=0$ and $A_3=0$, we will have $B_2=\partial A_1/\partial X_3$, and $B_3 = -\partial A_1/\partial X_2$. The $A_1$ component obeys the following equation
\beq
\left(\frac{\partial}{\partial t}+SX_1\frac{\partial}{\partial X_2}\right)A_1+u_2\frac{\partial A_1}{\partial X_2} + u_3\frac{\partial A_1}{\partial X_3} =\;\eta\nabla^2 A_1\,.
\eeq 
 The governing equation for $A_1$ is similar to the \Eq{b1_eqn} and therefore, $A_1$ component must also decay with time. If there is no growth in the single realisation of the field, there will not be growth even in the ensemble averaged mean-field. The strong shear makes the flow quasi-two dimensional and hence magnetic field decays in such a flow. One can workout in detail the nature of the decay and the rate of the decay as given in \cite{Kol17} for the above mentioned flow with shear background. It would be an interesting exercise to carry out such an analysis.

\section{Averaging method}
\label{Aveg_tech}
Here we outline the averaging procedure that we have used both in this work and in paper~I. The averaging over the phase $\Psi$ is performed analytically (see paper~I and \cite{KSS12}). The averaging over the triad $(\bfq,\bfa,\bfc)$ is performed numerically. We need to average $\bfq$ over the sphere of radius $q$. We will express the vectors $(\bfq,\bfa,\bfc)$ in terms of magnetic field wavevector $\bfk$ by the Euler angles.  
Let $\bfq$ make an angle $\theta$ with the direction of $k_3$ and $\phi$ be the angle between projection of $\bfq$ on $k_1-k_2$ plane and the direction of $k_1$. The angle $\psi$ is in the plane formed by the vectors $\bfa$ and $\bfc$. To average $\bfq$ over the sphere of radius $q$, $\theta$ needs to vary from $0$ to $\pi$, $\phi$ needs to vary from $0$ to $2\pi$. To average $\bfa$ over the circle perpendicular to the vector $\bfq$, $\psi$ has to vary from $0$ to $2\pi$. Using the three Euler angles for the rigid body rotation, we can relate the triad $(\bfq,\bfa,\bfc)$ to the direction of  magnetic field wave vector $(k_1,k_2,k_3)$ at time zero as:

\begin{figure}
\centering
\scalebox{1} 
{
\begin{pspicture}(1.0,-4.5)(9.09,4.5)
\psline[linewidth=0.04cm,arrowsize=0.05291667cm 2.0,arrowlength=1.4,arrowinset=0.4]{->}(3.21,-1.62)(3.23,3.96)
\psline[linewidth=0.04cm,arrowsize=0.05291667cm 2.0,arrowlength=1.4,arrowinset=0.4]{->}(3.23,-1.64)(9.07,-1.66)
\psline[linewidth=0.04cm,arrowsize=0.05291667cm 2.0,arrowlength=1.4,arrowinset=0.4]{->}(3.21,-1.62)(0.35,-3.96)
\psline[linewidth=0.04cm,arrowsize=0.05291667cm 2.0,arrowlength=1.4,arrowinset=0.4]{->}(3.23,-1.62)(5.93,2.22)
\psline[linewidth=0.04cm,linestyle=dotted,dotsep=0.26cm](5.87,2.14)(5.83,-3.0)
\psline[linewidth=0.04cm,linestyle=dashed,dash=0.16cm 0.16cm](5.81,-3.08)(1.47,-3.06)
\psline[linewidth=0.04cm,linestyle=dashed,dash=0.16cm 0.16cm](5.83,-3.08)(7.27,-1.66)
\psline[linewidth=0.02cm,linestyle=dashed,dash=0.16cm 0.16cm](3.21,-1.66)(5.83,-3.04)
\psarc[linewidth=0.03](3.51,-0.96){0.32}{21.801409}{146.30994}
\psarc[linewidth=0.03](3.19,-1.72){0.34}{207.89728}{339.44397}
\usefont{T1}{ptm}{m}{n}
\rput(3.6845312,-0.33){$\theta$}
\usefont{T1}{ptm}{m}{n}
\rput(3.2545311,-2.35){$\phi$}
\usefont{T1}{ptm}{m}{n}
\rput(0.46453124,-3.29){$k_1$}
\usefont{T1}{ptm}{m}{n}
\rput(8.484531,-1.37){$k_2$}
\usefont{T1}{ptm}{m}{n}
\rput(3.6445312,3.49){$k_3$}
\usefont{T1}{ptm}{m}{n}
\rput(5.324531,1.93){$\bfq$}
\psline[linewidth=0.03cm,arrowsize=0.05291667cm 2.0,arrowlength=1.4,arrowinset=0.4]{->}(5.21,1.2)(5.63,0.2)
\psline[linewidth=0.03cm,arrowsize=0.05291667cm 2.0,arrowlength=1.4,arrowinset=0.4]{->}(5.21,1.2)(6.13,1.46)
\psline[linewidth=0.03cm,linestyle=dashed,dash=0.16cm 0.16cm](5.23,1.2)(5.83,0.84)
\psarc[linewidth=0.03](5.38,0.99){0.13}{258.69006}{18.434948}
\usefont{T1}{ptm}{m}{n}
\rput(5.704531,0.62){$\psi$}
\usefont{T1}{ptm}{m}{n}
\rput(5.2545315,0.35){$\bfa$}
\usefont{T1}{ptm}{m}{n}
\rput(5.6845314,1.53){$\bfc$}
\end{pspicture} 
}
\end{figure}
\beq
\begin{pmatrix}
\hat{\bfa} \\
\hat{\bfc} \\
\hat{\bfq} 
\end{pmatrix}
= 
\begin{pmatrix}
\cos{\theta}\cos{\psi}\cos{\phi}-\sin{\phi}\sin{\psi} & \cos{\theta}\cos{\psi}\sin{\phi}+\cos{\phi}\sin{\psi} & -\cos{\psi}\sin{\phi} \\
-\cos{\theta}\cos{\phi}\sin{\psi}-\sin{\phi}\cos{\psi} & -\cos{\theta}\sin{\phi}\sin{\psi}+\cos{\phi}\cos{\psi} & \sin{\phi}\sin{\psi} \\
\sin{\theta}\cos{\phi} & \sin{\theta}\sin{\phi} & \cos{\theta}
\end{pmatrix}
\begin{pmatrix}
\hat{k}_1 \\
\hat{k}_2 \\
\hat{k}_3 \\
\end{pmatrix}
\eeq
Since the values of cosines and sines vary in the range $[-1,1]$, we can employ the following transformation for averaging
\beq
Z = \cos\theta, \quad Y = \cos\phi, \quad X = \cos\psi
\label{trans}
\eeq
A general function $F(\theta,\phi,\psi)$ is averaged over the sphere of radius $q$ and the over the circle with radius $|\bfa|$ perpendicular to the direction of $\bfq$ as follows,

\beq
\displaystyle\int_0^{2\pi}\int_0^{2\pi}\int_0^{\pi} F(\theta,\phi,\psi) \frac{\sin\theta\,\ud\theta}{2}\frac{\ud\phi}{2\pi}\frac{\ud \psi}{2\pi}
\label{int_trig}
\eeq
once we employ the transformation given in \Eq{trans}, we get

\begin{align}
&\frac{1}{8\pi^2}\int_{-1}\ud Z\int_{-1}^1\frac{2\,\ud X}{\sqrt{1-X^2}}\int_{-1}^1\frac{2\ud Y}{\sqrt{1-Y^2}}{\cal F}(X,Y,Z) \nonumber \\
& \approx \displaystyle\sum_{i,j,k}^{N}w'_i\,w_j\,w_k\,{\cal F}_{ijk}  
\label{int_xyz}
\end{align}
where ${\cal F}(X,Y,Z)=F(\theta,\phi,\psi)$. We use Gauss quadrature rule to integrate the functions. We can use Gauss Chebyshev quadrature to integrate over $X$ and $Y$, since we can use $1/\sqrt{1-X^2}$ and $1/\sqrt{1-Y^2}$ as a weight functions for integration. And we use Gauss Legendre quadrature to integrate over variable $Z$. In \Eq{int_xyz} $w$ and $w'$ are the weights corresponding to Chebyshev and Legendre polynomials, respectively; and ${\cal F}_{ijk}$ is evaluated at the zeros of the Chebyshev and Legendre polynomials. As the $N$ value in the summation increases, the approximation becomes better. We use $N=80$, which gives accuracy upto fourth decimal place.  Evaluating \Eq{int_xyz} numerically is much faster as compared to evaluation of \Eq{int_trig} as trigonometric function evaluation is slower as compared to the evaluation of polynomials and square roots.

\label{lastpage}

\end{document}